\documentclass[twocolumn,trackchanges]{aastex701}

\usepackage{graphicx}
\usepackage{amsmath}
\usepackage{url}
\usepackage{enumerate}
\usepackage{multirow}
\usepackage{lineno}
\usepackage{soul}
\usepackage{bm}
\usepackage{float}
\usepackage{soul}
\usepackage{tablefootnote}
\usepackage{booktabs}
\usepackage{longtable}
\linenumbers

\hypersetup{colorlinks = true, linkcolor = green, anchorcolor = red, citecolor = blue, filecolor = red, pagecolor = red, urlcolor = red}
\usepackage{url}
\usepackage{colortbl}

\RequirePackage{color}

\newfont{\myfont}{cmmib10}

\def\lapprox{\mathrel{\hbox{\rlap{\hbox{\lower4pt\hbox{$\sim$}}}\hbox{$<$}}}}
\def\gapprox{\mathrel{\hbox{\rlap{\hbox{\lower3pt\hbox{$\sim$}}}\hbox{$>$}}}}

\definecolor{apricot}{rgb}{0.88,0.81,0.65}

\graphicspath{{./}{figures/}}

\begin{document}

\correspondingauthor{Habtamu M. Tedila}
\email{habtamu.tedila@mcgill.ca}

\author[orcid=0000-0002-5815-6548,gname=Habtamu,sname=Tedila]{H.~M. Tedila}
\affiliation{Trottier Space Institute, McGill University, 3550 rue University, Montréal, QC H3A 2A7, Canada}
\affiliation{Department of Physics, McGill University, 3600 rue University, Montréal, QC H3A 2T8, Canada}
\email{habtamu.tedila@mcgill.ca}

\author[gname=Shu-Jin,sname=Dang]{S.~J. Dang}
\affiliation{School of Physics and Electronic Science, Guizhou Normal University, Guiyang, 550001, People’s Republic of China}
\email{dangsj@gznu.edu.cn}

\author[orcid=0000-0003-3010-7661,gname=Di,sname=Li]{D. Li}
\affiliation{New Cornerstone Science Laboratory, Department of Astronomy, Tsinghua University, Beijing, People's Republic of China}
\affiliation{National Astronomical Observatories, Chinese Academy of Sciences, A20 Datun Road, Chaoyang District, Beijing 100101, People's Republic of China}
\affiliation{State Key Laboratory of Radio Astronomy and Technology, Beijing 100101, People's Republic of China}
\email{dili@tsinghua.edu.cn}

\author[orcid=0000-0002-3386-7159,gname=Pei,sname=Wang]{P. Wang}
\affiliation{National Astronomical Observatories, Chinese Academy of Sciences, A20 Datun Road, Chaoyang District, Beijing 100101, People's Republic of China}
\affiliation{Institute for Frontiers in Astronomy and Astrophysics, Beijing Normal University, Beijing 102206, People's Republic of China}
\affiliation{State Key Laboratory of Radio Astronomy and Technology, Beijing 100101, People's Republic of China}
\email{wangpei@nao.cas.cn}

\author[orcid=0000-0002-5381-6498,gname=Jian-Ping,sname=Yuan]{J.~P. Yuan}
\affiliation{Xinjiang Astronomical Observatory, Chinese Academy of Sciences, 150 Science 1-Street, Urumqi, Xinjiang, 830011, People's Republic of China}
\affiliation{Xinjiang Key Laboratory of Radio Astrophysics, 150 Science1-Street, Urumqi, Xinjiang, 830011, People's Republic of China}
\email{yuanjp@xao.ac.cn}

\author[gname=Rai,sname=Yuen]{R. Yuen}
\affiliation{Xinjiang Astronomical Observatory, Chinese Academy of Sciences, 150 Science 1-Street, Urumqi, Xinjiang, 830011, People's Republic of China}
\affiliation{Xinjiang Key Laboratory of Radio Astrophysics, 150 Science1-Street, Urumqi, Xinjiang, 830011, People's Republic of China}
\email{ryuen@xao.ac.cn}

\author[orcid=0000-0002-9786-8548,gname=Na,sname=Wang]{N. Wang}
\affiliation{Xinjiang Astronomical Observatory, Chinese Academy of Sciences, 150 Science 1-Street, Urumqi, Xinjiang, 830011, People's Republic of China}
\affiliation{Xinjiang Key Laboratory of Radio Astrophysics, 150 Science1-Street, Urumqi, Xinjiang, 830011, People's Republic of China}
\email{na.wang@xao.ac.cn}

\author[gname=Shakhboz,sname=Khasanov]{Sh. Khasanov}
\affiliation{Institute of Nuclear Technologies, Samarkand State University}
\email{shakhbozkhasanov93@gmail.com}

	
\title{Timing, Polarization, and Single-Pulse Properties of Long-Period FAST Pulsars}

\begin{abstract}
 
We present phase-connected timing and polarization measurements for two long-period FAST-CRAFTS pulsars, PSRs J0000+6252 and J2131+3642, and extend single-pulse emission-state analysis to a five-source sample including PSRs J1903+1407, J1502+4653, and J2112+4058.
FAST L-band timing baselines span 414 to 511 days; the two pulsars have spin periods of 1.11 to 1.55 s, period derivatives of $(3.99$ to $4.06)\times10^{-15}~{\rm s~s^{-1}}$, characteristic ages of 4.34 to 6.16 Myr, surface magnetic fields of $(2.15$ to $2.52)\times10^{12}$ G, and spin-down luminosities of $(4.21\times10^{31}$ to $1.16\times10^{32})~{\rm erg~s^{-1}}$.
Their rotation measures are $70.2\pm26.7$ and $-44.6\pm7.6~{\rm rad~m^{-2}}$, with linear polarization fractions of 17.6\% to 27.2\%.
PSR J2131+3642 shows a short monotonic PA segment permitting a formal rotating-vector-model fit, though limited longitude coverage leaves the geometric parameters poorly constrained; PSR J0000+6252 has too few PA points for such a fit.
Gaussian mixture modeling (GMM) of single-pulse energy distributions identifies null, weak, and burst components in PSRs J0000+6252, J1903+1407, J1502+4653, and J2112+4058, while J2131+3642 shows only weak and burst states.
We also identify bright single pulses (peak intensity $\geq10\times$ the integrated average profile): 42 in J0000+6252, four each in J1903+1407 and J2112+4058, and none in J2131+3642 or J1502+4653.
These bright pulses occur within the main emission window with no evidence of periodic recurrence, consistent with sporadic enhancements of the normal radio-emission beam.
For J2112+4058, we measure a scattering timescale $\tau_{\rm sc}=5.84\pm0.18$ ms at $\nu_{\rm ref}=1.25$ GHz.
Together, these results highlight the diversity of magnetospheric variability among slowly rotating neutron stars.
 
\end{abstract}

	
\setcounter{footnote}{0}

\section{INTRODUCTION}\label{sec:intro}
	
Long-term monitoring of pulsar pulse arrival times measures spin period, spin-down rate, astrometric position, dispersion measure, and, when applicable, binary motion \citep{Weisberg2010}. 
These measurements provide characteristic ages, surface magnetic fields, spin-down luminosities, and distance estimates \citep{Becker2018}. 
Long-period pulsars are especially valuable because they probe coherent radio emission in slowly rotating neutron stars, where pair production and magnetospheric plasma supply may approach their operating limits \citep{Rea2024, Tedila2026}. 
They also connect timing behavior, magnetospheric activity, and Galactic population studies \citep{Cordes1990, LagunaWolszczan1997, Zhu2015, Kramer2024}. 
Discovering and following up long-period pulsars with the Five-hundred-meter Aperture Spherical radio Telescope (FAST) therefore provides an important route to understanding how radio emission persists late in neutron-star evolution \citep{Nan13, Li18, Wu2023, Tedila2025}.

While pulsars serve as stable clocks, they also display timing noise, glitches, slow glitches, and spin-down changes \citep{Yuan2010, Shabanova2007}. 
These variations provide insights into neutron-star interiors and magnetospheric torques. Here, a magnetospheric state denotes a quasi-stable configuration of the plasma density, currents, and accelerating electric fields in the pulsar magnetosphere. 
Transitions between such configurations can alter both the radio-emission pattern and the torque responsible for spin-down. Spin-down changes often coincide with variations in radio emission, linking timing noise to changes in magnetospheric state \citep{Lyne2010}. 
Additional phenomena, including mode changes, nulling, intermittency, and subpulse drifting, further clarify the relationship between rotation and emission \citep{Rankin1986, Perera2015, BasuMitra2018, Tedila2022, Cai2025}. 
Expanding the sample of pulsars with measured rotation and emission is crucial for understanding magnetospheric evolution and timing stability \citep{Lyne2010, Wright2022}.

FAST, equipped with a 19-beam L-band receiver, has discovered over 1,000 pulsars, significantly enhancing survey capabilities \citep[e.g.,][]{Li18, Pan2021, Han2025}\footnote{\url{https://fast.bao.ac.cn/cms/article/32}}.
The Commensal Radio Astronomy FAST Survey (CRAFTS)\footnote{\url{https://crafts.bao.ac.cn/pulsar/}} has rapidly expanded the pulsar sample, identifying at least 74 millisecond pulsars, approximately 141 non-recycled pulsars, and seven confirmed rotating radio transients (RRATs) \citep{Miao2023}.
This extensive Galactic coverage enables novel tests of pulsar emission and evolutionary models \citep{Zhang2019}.
The Galactic Plane Pulsar Snapshot survey (GPPS)\footnote{\url{http://zmtt.bao.ac.cn/GPPS/}} has reported a large and diverse pulsar sample \citep{Han2025}, thereby expanding the known pulsar parameter space. The FAST Globular Cluster Pulsar Survey\footnote{\url{https://fast.bao.ac.cn/cms/article/65/}} has identified several new pulsars within globular clusters \citep{Pan2021}.
Subsequent timing observations have enhanced positional accuracy, verified rotational stability, determined spin-down rates, and generated phase-connected ephemerides, as demonstrated by early FAST results \citep{Zhao2024}.
Analyses by \citet{Cameron2020} and others have identified sources exhibiting distinctive emission and polarization profiles. 
Additional observations with the Effelsberg and Arecibo telescopes have underscored the diversity of emission among FAST discoveries \citep{Cruces2021, Wang2021R, Miao2023}.
Recent studies have presented timing and polarization measurements for pulsars discovered by FAST, revealing phenomena such as nulling and drifting, and refining models of pulsar radiation physics and Galactic electron density \citep{Cruces2021, Miao2023, Wu2023, Zhao2024, Yang2025}.

Millisecond pulsars support array timing, while long-period and intermittent sources test emission models and death-line predictions \citep{D'Amico1998, Ghosh2025}. 
Green Bank Telescope follow-up of FAST-CRAFTS pulsars has refined Galactic electron-density models when distance predictions differ \citep{Blackmon2026}.
Long-period pulsars, located in specific $P$--$\dot{P}$ regions, clarify the conditions for sustained radio emission \citep{Jones2022, Tedila2022, Tedila2025}. 
Phenomena such as nulling, intermittency, or drifting indicate magnetospheres near emission thresholds, and pulsars near the death line test pair-production and current-emission models \citep{Ruderman1975, Chen1993, Zhang2000}. 
Accurate dispersion measures (DMs) from these pulsars further constrain the Galactic electron distribution, especially when NE2001 \citep{NE2001}, YMW16 \citep{YMW16}, and NE2025 \citep{NE2025} estimates differ \citep{Price2021}.

In this work, we present phase-connected timing solutions and polarization measurements for PSRs J0000+6252 and J2131+3642. We include PSRs J1903+1407, J1502+4653, and J2112+4058 only in the single-pulse emission-state analysis. 
Together, these five sources have spin periods between 1.11 and 4.06~s and sample a region of parameter space where radio-emission stability is expected to be sensitive to magnetospheric state.  
By combining timing, polarization, and pulse-energy statistics, we add two timing solutions and characterize, for each of five pulsars, the fractions and contiguous durations of null, weak, and burst emission states inferred from pulse-energy distributions.

\section{Observations and Data Processing}\label{sec:obs}

FAST was completed in September 2016 and has a maximum effective aperture of approximately 300 m \citep{Li18, Jiang19}. 
The two pulsars analyzed in this study were selected for follow-up timing because they are a well-observed subset of recent long-period discoveries. 
For the two timed pulsars, multi-epoch coverage over more than one year enables phase-connected timing solutions and polarimetric characterization. We additionally include PSRs J1903+1407, J1502+4653, and J2112+4058 only in the single-pulse analyses; we do not present timing or polarization results for these three sources here.

Observations were conducted using the FAST 19-beam receiver across 1.05-1.45\,GHz \citep{Jiang19}. 
Data were recorded in 8-bit search-mode PSRFITS format \citep{Hotan04}, with 1024 frequency channels, a time resolution of 49.152 $\mu s$, and a frequency resolution of 0.488 MHz.

Timing and polarimetric follow-up observations of PSRs~J0000+6252 and J2131+3642 were conducted approximately monthly. Each session comprised 40~s of noise-diode injection for polarimetric calibration, 20~s for antenna slewing, and 240~s of on-source integration, for a total duration of approximately 300~s. During the noise-diode observation, the telescope was pointed \(10'\) away from the pulsar position. 
Table~\ref{tab:timing_observations} summarizes this multi-epoch data set. Each pulsar was observed in more than 12 sessions over a baseline exceeding one year.

The single-pulse analysis used one uninterrupted on-source sequence for each pulsar, acquired separately from the individual 240-s timing scans. The on-source durations were 2400~s for PSR~J0000+6252 and PSR~J1903+1407, 2220~s for PSR~J2131+3642, 3600~s for PSR~J1502+4653, and 4800~s for PSR~J2112+4058. These observations contain 1547, 1868, 1996, 2057, and 1182 rotations, respectively. 
Before calculating pulse energies and fitting the Gaussian mixture modeling (GMMs), rotations affected by radio-frequency interference were masked and excluded from the relevant calculations. Thus, the 240-s value refers to each short timing epoch, whereas the 2220--4800-s values refer to the dedicated, uninterrupted single-pulse sequences.

\begin{table}
\centering
\setlength{\tabcolsep}{5pt}
\renewcommand{\arraystretch}{1.25}
\caption{Summary of the timing data set for the two FAST-discovered long-period pulsars. The table lists the first and last observing epochs, the total timing baseline, and the number of TOAs used in the timing analysis.}
\label{tab:timing_observations}
\begin{tabular}{lcccc}
\hline
\hline
Pulsar & First epoch & Last epoch & Span & TOAs \\
J name & MJD & MJD & days & \\
\hline
J0000+6252 & 60273 & 60784 & 511 & 14 \\
J2131+3642 & 60421 & 60835 & 414 & 12 \\
\hline
\end{tabular}
\end{table}

Because these pulsars were newly discovered and lacked precise timing ephemerides, their initial $P$ and DMs were confirmed using PRESTO \citep{Ransom2011}. 
Initial positions, spin frequencies, and DMs were obtained from the CRAFTS survey \citep{Li18}. 
Periodic signals were identified through Fourier-transform searches of time-domain data from each beam. 
Candidate signals were selected based on period, and DM ranges typical of known pulsars. 
Accidental radio-frequency interference was excluded by verifying detection consistency across adjacent beams in the FAST 19-beam receiver.

The preliminary spin periods and DMs were used to fold the raw PSRFITS data with \texttt{DSPSR} \citep{Straten11}, producing an average pulse profile with 1024 phase bins for each observation. 
Strong narrowband and broadband radio-frequency interference (RFI) was identified and excised using the \texttt{paz} and interactive \texttt{pazi} tools in PSRCHIVE \citep{Hotan04}.
For each observation, integrated pulse profiles were obtained by summing over time, frequency, and polarization. After aligning the pulse phases of all observations, 
a high-signal-to-noise-ratio standard template was generated using \texttt{paas} by fitting Gaussian components to the integrated profile. 
This standard template was then used to calculate pulse times of arrival (TOAs) with the \texttt{pat} tool, which were subsequently used to refine the timing solutions.

\begin{table}
\centering
\setlength{\tabcolsep}{6pt}
\renewcommand{\arraystretch}{1.25}
\caption{Integrated-profile signal-to-noise ratios, pulse widths at 10\% and 50\% of the profile peak, and mean flux densities at 1.25~GHz, measured from the dedicated long-duration single-pulse observations.}
\label{tab:pulse_width_flux}
\begin{tabular}{lcccc}
\hline
\hline
Pulsar & S/N & $W_{10}$ & $W_{50}$  & $S_{\rm mean}$  \\
J name &  & (deg.) & (deg.) & ($\mu$Jy) \\
\hline
J1903+1407 & $105.1$ & $20.60(33)$ & $4.31(9)$ & $27.8(23)$ \\
J0000+6252 & $55.4$ & $9.66(23)$ & $4.03(18)$ & $9.9(11)$ \\
J2131+3642 & $121.2$ & $15.76(29)$ & $6.62(11)$ & $39.2(30)$ \\
J1502+4653 & $659.6$ & $9.63(9)$ & $2.22(3)$ & $92.9(23)$ \\
J2112+4058 & $2805.9$ & $8.24(6)$ & $4.13(13)$ & $319.7(83)$ \\
\hline
\end{tabular}

\vspace{0.5em}
\footnotesize
Note. Values in parentheses are $1\sigma$ bootstrap uncertainties in the last quoted digits.
\end{table}

\begin{table*}
\centering
\setlength{\tabcolsep}{30pt}
\renewcommand{\arraystretch}{1.25}
\caption{Timing solutions and derived parameters for the two long-period pulsars. Values in parentheses indicate 1$\sigma$ uncertainties in the last quoted digits. The timing conventions are reported directly from the final parameter files.}
\label{tab:timing_derived_parameters}

\begin{tabular}{lcc}
\hline
\hline
\textbf{Parameter} 
& \textbf{PSR~J0000+6252} 
& \textbf{PSR~J2131+3642} \\
\hline

\multicolumn{3}{l}{\textit{Astrometric and spin parameters}} \\
Right ascension, $\alpha$ (J2000) 
& 00:00:56.83(1) 
& 21:31:46.64(7) \\

Declination, $\delta$ (J2000) 
& +62:52:09.565(98) 
& +36:42:58.2(5) \\

Spin frequency, $\nu$ (s$^{-1}$) 
& 0.6441672996015(6) 
& 0.898788312596(3) \\

Spin frequency derivative, $\dot{\nu}$ ($10^{-15}$\,s$^{-2}$) 
& $-1.6560(9)$ 
& $-3.279(28)$ \\

Dispersion measure, DM (pc cm$^{-3}$) 
& 164(2) 
& 100.3(3) \\

\hline
\multicolumn{3}{l}{\textit{Timing data and fit quality}} \\
Solar-system ephemeris
& DE405
& DE405 \\

Terrestrial-time/clock correction
& TT(TAI)
& TT(TAI) \\

Barycentric time units
& TDB
& TDB \\

Time ephemeris
& FB90
& FB90 \\

First TOA (MJD) 
& 60273.561 
& 60421.039 \\

Last TOA (MJD) 
& 60784.098 
& 60835.774 \\

Timing epoch (MJD) 
& 60528 
& 60598 \\

Number of TOAs 
& 14 
& 12 \\

Weighted RMS ($\mu$s) 
& 266.651 
& 284.312 \\

EFAC 
& 1.373 
& 1.420 \\

\hline
\multicolumn{3}{l}{\textit{Galactic position and distance estimates}} \\
Galactic longitude, $l$ ($^\circ$) 
& 117.2123 
& 84.301 \\

Galactic latitude, $b$ ($^\circ$) 
& 0.5596 
& $-10.764$ \\

NE2001 distance (kpc) & 5.180 & 5.840 \\ 
NE2025 distance (kpc) & 4.195 & 8.535 \\ 
YMW16 distance (kpc) & 3.192 & 12.133 \\

\hline
\multicolumn{3}{l}{\textit{Derived physical parameters}} \\
Spin period, $P$ (s) 
& 1.55239174763(1) 
& 1.112609038174(47) \\

Spin period derivative, $\dot{P}$ ($10^{-15}$ s s$^{-1}$) 
& 3.991(2) 
& 4.059(35) \\

Characteristic age, $\tau_c$ (Myr) 
& 6.163 
& 4.343 \\

Surface magnetic field, $B_{\rm surf}$ ($10^{12}$ G) 
& 2.52 
& 2.15 \\

Spin-down luminosity, $\dot{E}$ ($10^{30}$ erg s$^{-1}$) 
& 42.1 
& 116.3 \\

\hline
\hline
\end{tabular}
\end{table*}

Phase-connected timing solutions were derived using {\tt\string TEMPO} \citep{Nice2015} and {\tt\string TEMPO2} \citep{Hobbs06}, based on the spin-down model
\begin{equation}\label{eq:spin_phase}
\phi(t) = \phi_0 + \nu(t-t_0) + \frac{1}{2}\dot{\nu}(t-t_0)^2 + \frac{1}{6}\ddot{\nu}(t-t_0)^3 ,
\end{equation}
where $\phi(t)$ is the barycentric pulse phase, $\phi_0$ is the phase at the reference epoch $t_0$, $\nu$ is the spin frequency, and $\dot{\nu}$ and $\ddot{\nu}$ are the first and second derivatives of the spin frequency, respectively.
Polarimetric calibration was performed using the `\texttt{pac}' tool in PSRCHIVE, constructing a calibration database from noise-diode observations at each epoch. 
This calibration corrects for differential gain and phase between receiver polarization channels. 
The DM was measured by dividing the observing bandwidth into two frequency subbands and computing a TOA for each subband. 
After calibration, the `\texttt{rmfit}' tool was used to estimate the Faraday rotation measure (RM) for each pulsar. For both sources, the refined search used 401 trial RMs uniformly spanning $-200$ to $+200~{\rm rad\,m^{-2}}$, corresponding to a trial spacing of $1~{\rm rad\,m^{-2}}$.

Since our FAST data were not on an absolute flux scale, we estimated each pulsar’s mean flux density at 1.25 GHz using its integrated-profile signal-to-noise ratio and the standard radiometer equation \citep{Lorimer2005}:
\begin{equation}\label{eq:radiometer_mean}
S_{\rm mean} = \frac{\beta ({\rm S/N}) T_{\rm sys}} {G \sqrt{n_{\rm pol} b t_{\rm obs}}} \left(\frac{W_{10}}{P-W_{10}}\right)^{1/2},
\end{equation}
Here, $S_{\rm mean}$ is the estimated mean flux density, $\beta$ is the digitization-loss correction factor,
$S/N$ is the integrated signal-to-noise ratio of the folded average profile, measured as
$\sum I_{\rm on}/(\sigma_{\rm off}\sqrt{N_{\rm on}})$, $T_{\rm sys}$ is the system temperature, $G$ is the telescope gain, $n_{\rm pol}$ is the number of summed polarizations,
$b$ is the effective bandwidth, $t_{\rm obs}$ is the integration time, $P$ is the spin period, and $W_{10}$ is the pulse width at 10 percent of peak intensity.
We adopted $\beta=1$, $n_{\rm pol}=2$, $b=400~{\rm MHz}$, and $G=16.12~{\rm K~Jy^{-1}}$ for FAST at 1.25 GHz \citep{Jiang20}. 
The system temperature was estimated as a function of zenith angle using the empirical FAST receiver model from \citet{Jiang20}:
\begin{equation}
T_{\rm sys} = P_{0}\arctan\left(\sqrt{1+\theta_{\rm ZA}^{n}}-P_{1}\right)+P_{2},
\end{equation}
$\theta_{\rm ZA}$ is the zenith angle in degrees. For beam M01 at 1250 MHz, we used $P_0=4.35$, $P_1=6.88$, $P_2=25.54$, and $n=1.21$. 
Since these fluxes are based on the radiometer equation rather than absolute flux calibration, they are approximate.
The estimated mean flux densities are listed in Table~\ref{tab:pulse_width_flux}. 
Pulse widths were measured at 10\% and 50\% of the integrated profile peak, and we adopted $\theta_{\rm ZA}=20^\circ$. 
Quoted uncertainties are $1\sigma$ bootstrap errors, propagated to $S_{\rm mean}$ through the radiometer equation (Equation~\ref{eq:radiometer_mean}).

\begin{figure*}[ht]
    \centering
    \begin{minipage}{0.45\textwidth}
        \centering
        \textbf{(a) PSR~J0000+6252}\\
        \includegraphics[width=\linewidth]{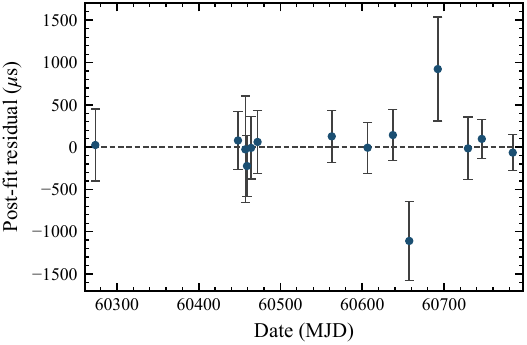}
    \end{minipage}
    \begin{minipage}{0.45\textwidth}
        \centering
        \textbf{(b) PSR~J2131+3642}\\
        \includegraphics[width=\linewidth]{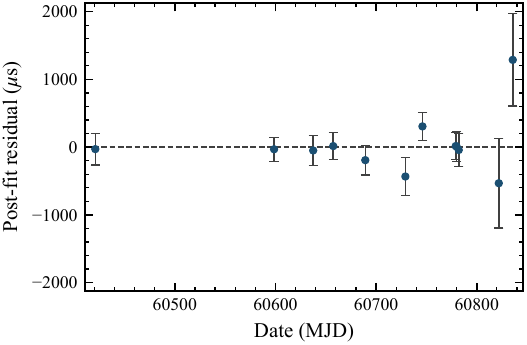}
    \end{minipage}
    \caption{Timing residuals for the two pulsars. Each panel shows the timing residuals in microseconds as a function of MJD for PSRs~J0000+6252 and J2131+3642, respectively. The dashed horizontal line at zero indicates the expected arrival times based on the timing model, 
    while the error bars represent the measurement uncertainties.}
    \label{fig:residuals}
\end{figure*} 

\begin{figure*}[ht]
    \centering
    \begin{minipage}{0.45\textwidth}
        \centering
        \textbf{(a) PSR~J0000+6252}\\
        \includegraphics[width=\linewidth]{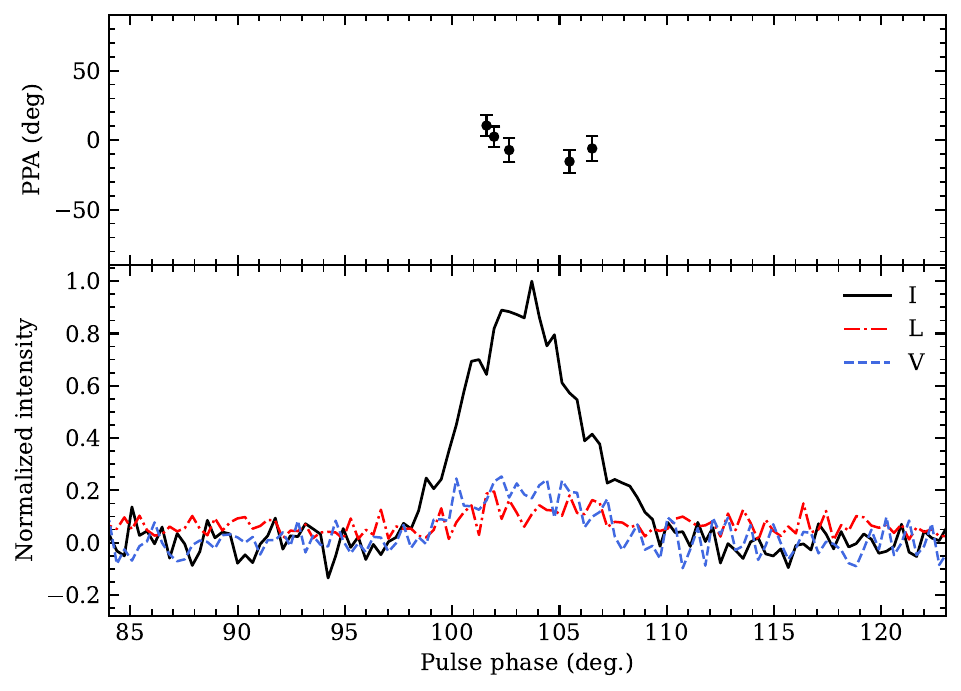}
    \end{minipage}
    \begin{minipage}{0.45\textwidth}
        \centering
        \textbf{(b) PSR~J2131+3642}\\
        \includegraphics[width=\linewidth]{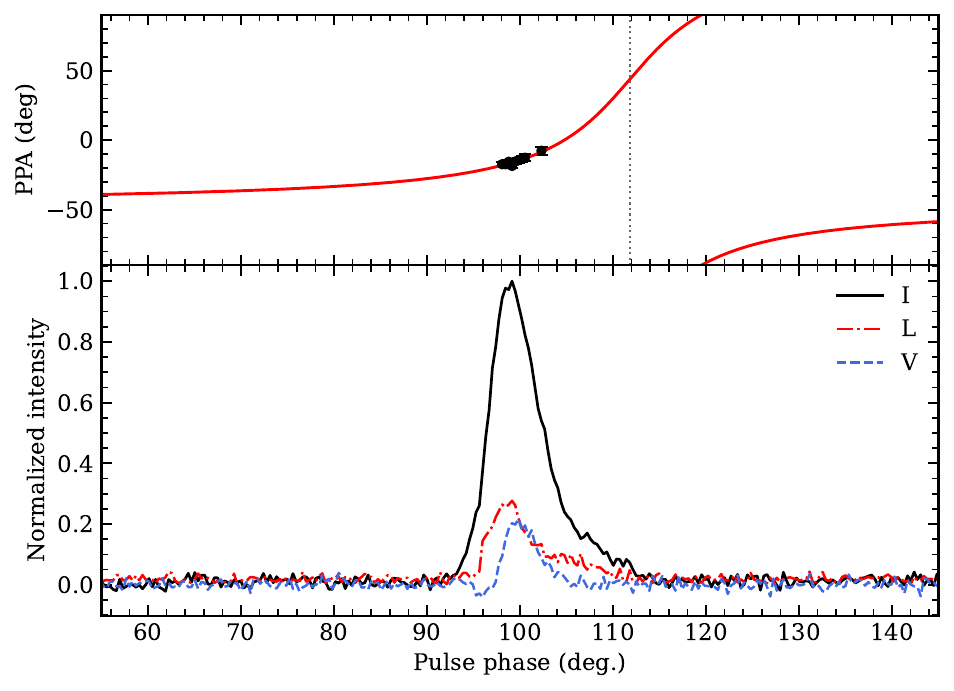}
    \end{minipage}
    \caption{Polarization profiles for (a) PSR~J0000+6252 and (b) PSR~J2131+3642. The lower sub-panels show normalized total intensity $I$ (black solid), linear polarization $L$ (red dash-dotted), and circular polarization $V$ (blue dashed). 
    Black points with $1\sigma$ error bars in the upper sub-panels are PA measurements retained only where the debiased linear-polarization signal-to-noise ratio exceeds 3; this selection suppresses noise-dominated and undefined PAs. 
    For PSR~J2131+3642, the red curve is the maximum-posterior rotating-vector-model fit, with $(\alpha,\beta,\phi_0,\psi_0)=(94.4^\circ,-7.4^\circ,111.8^\circ,43.7^\circ)$; the vertical dotted line marks $\phi_0$.}
    \label{fig:polarization_profiles}
\end{figure*}

\section{Timing Analysis}\label{sec:timing}

\subsection{Timing Solutions and Derived Parameters}

Phase-connected timing solutions were obtained for the two pulsars using standard techniques \citep{Nice2015, Hobbs06}. 
The timing residuals are shown in Figure~\ref{fig:residuals}, and the corresponding timing parameters and derived physical properties are summarized in Table~\ref{tab:timing_derived_parameters}. 
This analysis highlights the diversity and scientific value of long-period pulsars in the FAST sample \citep{Wu2023, Zhao2024}.

The measured spin periods span a range from 1.11 to 1.55 seconds, with period derivatives from $3.99\times10^{-15}$ to $4.06\times10^{-15}$ s s$^{-1}$. 
These values locate the pulsars in the long-period, ordinary-field regime of the $P$--$\dot{P}$ diagram, as shown in Figure~\ref{fig:P_Pdot}, and reinforce their classification as non-recycled, isolated neutron stars. 
The characteristic ages range from 4.34 to 6.16 Myr. Surface magnetic field strengths are estimated from $2.15\times10^{12}$ G to $2.52\times10^{12}$ G, and spin-down luminosities range from $4.21\times10^{31}$ erg s$^{-1}$ to $1.16\times10^{32}$ erg s$^{-1}$.

Timing residuals for each pulsar are well described by the models with no significant trends or large systematic deviations, indicating that rotational behavior is stable over the observed time span. 
The weighted RMS of the residuals (267--284 $\mu$s) and applied EFAC values show that time-of-arrival uncertainties are well accounted for in the analysis.
These results demonstrate the capability of FAST for precise long-term timing of faint, long-period pulsars.

\subsection{Distance Estimates and Model Discrepancies}

Distance estimates based on Galactic electron-density models, such as NE2001 \citep{NE2001}, NE2025 \citep{NE2025}, and YMW16 \citep{YMW16}, often exhibit significant discrepancies for the same pulsar, particularly at high Galactic latitudes or 
low DM \citep{Deller2019, YMW16}. These models differ in their treatment of Galactic structure, the inclusion of local features, and the underlying datasets used for calibration, 
leading to uncertainties of up to several factors in some cases \citep[e.g.,][]{Verbiest2012, Chatterjee2009}.

For example, the NE2001 and YMW16 distances for PSR~J2131+3642 differ by more than a factor of two, and for PSR~J0000+6252 by nearly a factor of two,
underscoring the need for independent distance constraints—such as annual parallax measurements \citep{Desvignes2016, Deller2019}, 
VLBI astrometry \citep{Chatterjee2009, Reardon2023}, or HI absorption studies \citep{Frail1990, Verbiest2012}—to refine Galactic electron-density models.

Accurate distance measurements are crucial for deriving intrinsic pulsar parameters, such as luminosity and velocity, 
and for understanding the spatial distribution and evolution of the Galactic pulsar population \citep{Lorimer2005, Manchester2005}. 
Recent studies have shown that errors in distance estimates can lead to systematic biases 
in neutron star population syntheses and models of Galactic structure \citep{Deller2019, YMW16}.
To address these discrepancies, we will conduct follow-up observations to obtain direct distance measurements via annual parallax and, 
where possible, HI absorption studies. This will yield more robust distance estimates, refine Galactic electron-density models, 
and enhance the scientific output of the FAST pulsar survey.

\begin{table}
\centering
\setlength{\tabcolsep}{8pt}
\renewcommand{\arraystretch}{1.25}
\caption{Polarization properties of the two pulsars with phase-connected timing solutions. The table lists the rotation measure (RM), linear polarization fraction ($L/I$), circular polarization fraction ($V/I$), and absolute circular polarization fraction ($|V|/I$). RM uncertainties are the $1\sigma$ values from \texttt{rmfit}.}
\label{tab:pulse_pol}
\begin{tabular}{lcccc}
\hline
\hline
Pulsar & RM & $L/I$ & $V/I$ & $|V|/I$ \\
 J name & (rad m$^{-2}$) & (\%) & (\%) & (\%) \\
\hline
J0000+6252 & $70.2\pm26.7$ & $17.6$ & $25.4$ & $25.4$ \\
J2131+3642 & $-44.6\pm7.6$ & $27.2$ & $12.9$ & $14.3$ \\
\hline
\end{tabular}

\vspace{0.5em}
\footnotesize
\end{table}

\subsection{Polarization Properties and Emission Geometry}

Table~\ref{tab:pulse_pol} summarizes the polarization properties of the two pulsars with phase-connected timing solutions. Their RM estimates are $70.2\pm26.7$ and $-44.6\pm7.6$~rad~m$^{-2}$ for PSRs~J0000+6252 and J2131+3642, respectively. 
Their linear polarization fractions ($L/I$) range from 17.6\% to 27.2\%. Signed circular polarization fractions ($V/I$) range from 12.9\% to 25.4\%, while absolute circular polarization fractions ($|V|/I$) range from 14.3\% to 25.4\%.

\begin{figure*}
    \centering
    \includegraphics[width=0.95\textwidth]{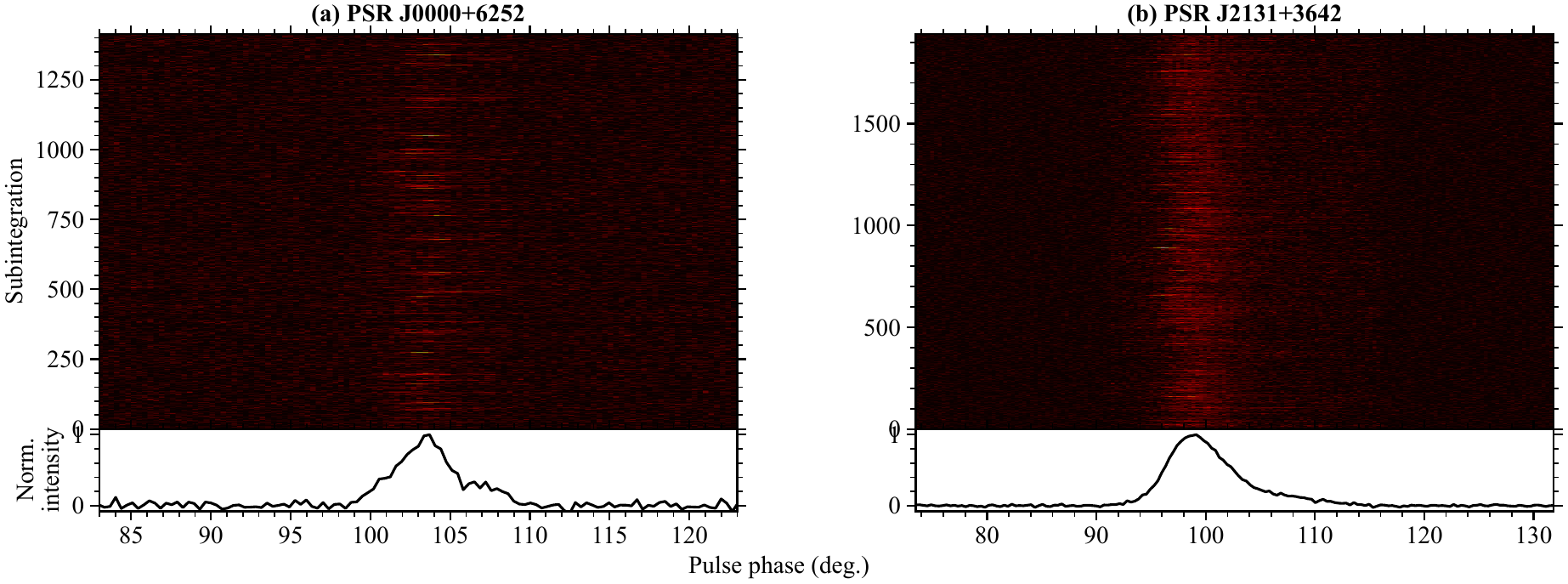}
    \caption{Single-pulse stacks and integrated profiles for (a) PSR~J0000+6252 and (b) PSR~J2131+3642, the two sources with timing and polarization measurements in this work. In each panel, the upper sub-panel shows intensity as a function of pulse number and rotational phase, and the lower sub-panel shows the corresponding normalized integrated profile.}
    \label{fig:pulse_stack_timed}
\end{figure*}

\begin{figure*}
    \centering
    \includegraphics[width=0.95\textwidth]{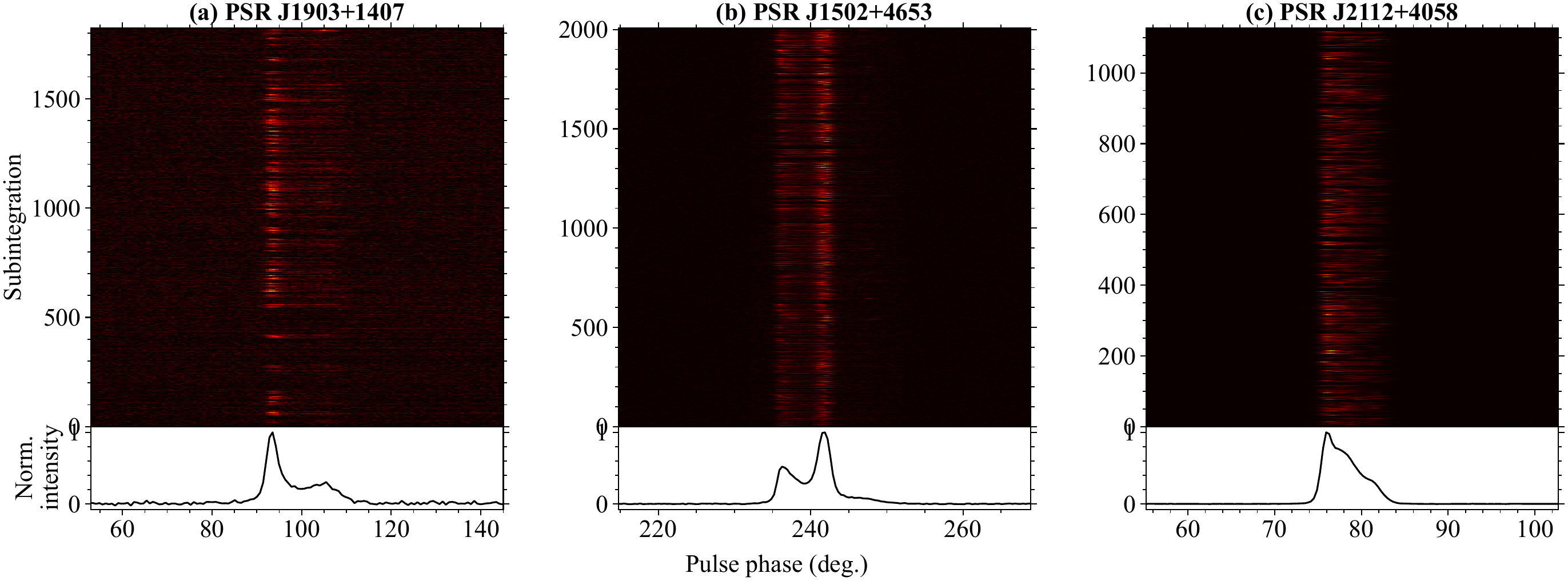}
    \caption{Similar to Figure~\ref{fig:pulse_stack_timed}, but for  (a) PSR~J1903+1407, (b) PSR~J1502+4653, and (c) PSR~J2112+4058.}
    \label{fig:pulse_stack_additional}
\end{figure*}

The observed polarization characteristics provide insight into the magnetospheric geometry and emission mechanisms of each pulsar \citep{Wright2022}. 
A widely used tool for interpreting polarization position-angle (PA) sweeps is the rotating vector model (RVM; \citealt{Radhakrishnan69}, \citealt{EverettWeisberg2001}), 
which relates the observed PA to the geometry of the pulsar's magnetic and rotational axes:
\begin{equation}
\psi(\phi) = \psi_0 + \arctan\left( \frac{\sin\alpha\, \sin(\phi-\phi_0)}{\sin\zeta\, \cos\alpha - \cos\zeta\, \sin\alpha\, \cos(\phi-\phi_0)} \right)
\end{equation}
Here, $\psi(\phi)$ is the observed linear polarization PA at pulse longitude $\phi$, $\psi_0$ is the PA at the closest approach, $\alpha$ is the angle between the rotation and magnetic axes, 
$\zeta=\alpha+\beta$ is the angle between the rotation axis and the observer's line of sight, and $\phi_0$ is the phase of the closest approach to the magnetic axis.

Figure~\ref{fig:polarization_profiles} shows that PSR~J0000+6252 has too few reliable PA points for RVM fitting. For PSR~J2131+3642, we fitted the monotonic segment defined by nine PA measurements that satisfy the linear-polarization signal-to-noise threshold described in the caption. 
The maximum-posterior solution is $(\alpha,\beta,\phi_0,\psi_0)=(94.4^\circ,-7.4^\circ,111.8^\circ,43.7^\circ)$, with $\chi^2_\nu=0.83$. 
However, the limited PA coverage produces broad, strongly covariant constraints: $\alpha=60.8^{+58.5}_{-35.6}$~deg, $\beta=-16.1^{+24.1}_{-27.0}$~deg, $\phi_0=94.2^{+50.9}_{-48.0}$~deg, and $\psi_0=-7.5^{+67.2}_{-51.7}$~deg (16th--84th percentiles). 
The red curve is therefore a formal fit to the observed PA segment, not a robust measurement of the viewing geometry.

In principle, RM probes the strength and direction of the line-of-sight magnetic field in the interstellar medium \citep{Han2017}. PSR~J2131+3642 provides a significant RM measurement, whereas the PSR~J0000+6252 value remains marginal; consequently, we do not draw a strong source-to-source comparison from the two RM values alone. 
The integrated profiles and polarization morphology of both sources resemble those of ordinary radio pulsars. The measured polarization fractions by themselves do not uniquely diagnose magnetospheric structure, because they can also depend on viewing geometry, orthogonal-mode mixing, propagation, and signal-to-noise ratio. 
We therefore report the source-to-source differences descriptively and do not interpret them as evidence for different magnetospheric structures or emission processes. Higher-S/N, wider-longitude, and multi-frequency polarimetry would be required for such an inference.

\begin{figure*}[ht]
    \centering
    \includegraphics[width=0.95\textwidth]{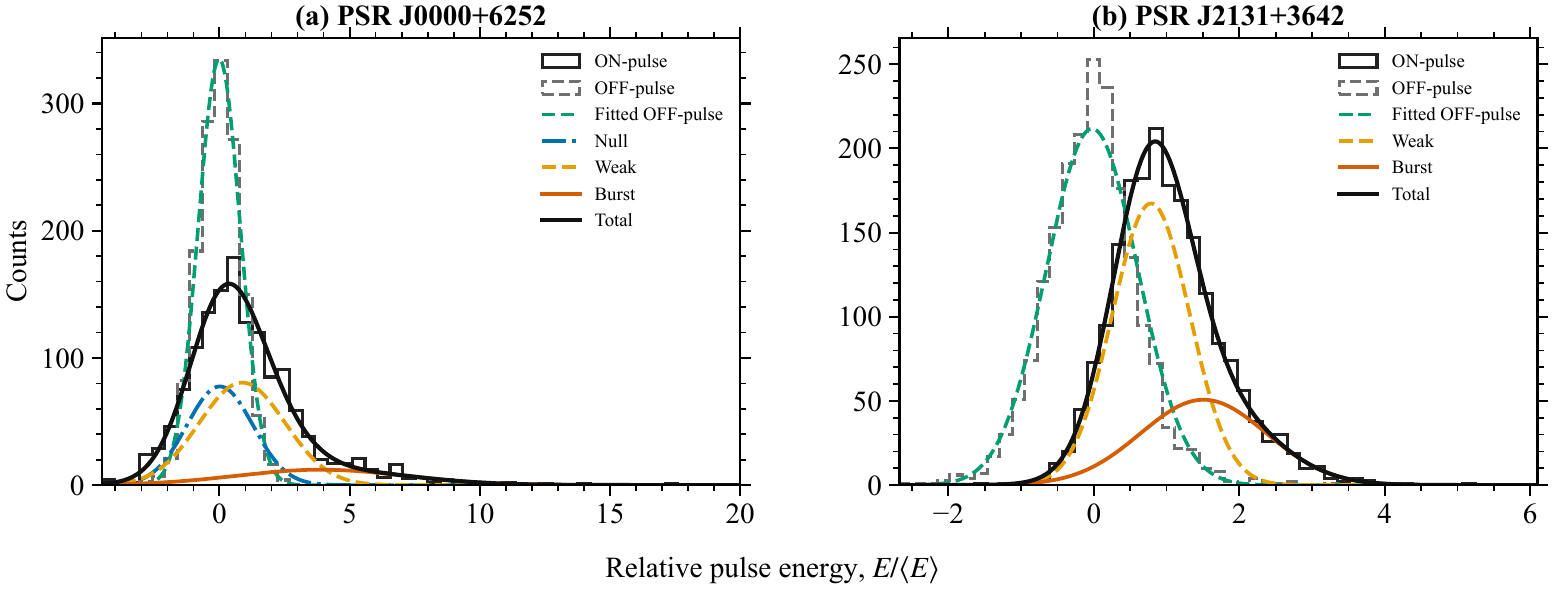}
    \caption{Normalized on-pulse and off-pulse energy distributions for (a) PSR~J0000+6252 and (b) PSR~J2131+3642. The histograms show the measured distributions, and the black curve shows the total Gaussian mixture model. The colored curves show the fitted off-pulse noise and the null, weak, and burst emission states for PSR~J0000+6252; PSR~J2131+3642 has only weak and burst emission states.}
    \label{fig:energy_timed}
\end{figure*}

\begin{figure*}[ht]
    \centering
    \includegraphics[width=0.95\textwidth]{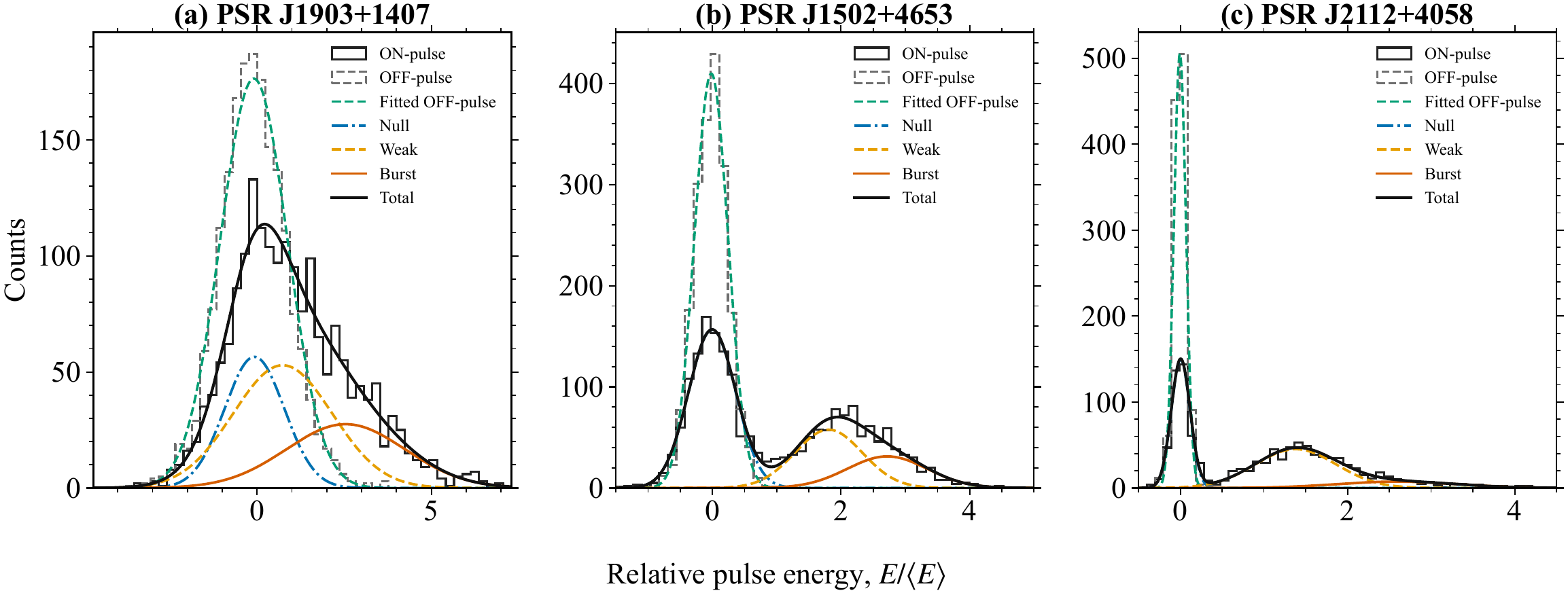}
    \caption{Similar to Figure~\ref{fig:energy_timed}, but for (a) PSR~J1903+1407, (b) PSR~J1502+4653, and (c) PSR~J2112+4058.}
    \label{fig:energy_additional}
\end{figure*}

\section{Single-Pulse Properties Analysis}\label{sec:single_pulse}

Single-pulse analysis probes emission variability that is hidden in integrated profiles. In long-period pulsars, nulling, mode changing, and strong pulse-to-pulse modulation are common \citep[e.g.,][]{Tedila2022, Grover2024, Tedila2024, Tedila2025}. 
These phenomena help distinguish persistent emitters from intermittent or frequently nulling sources and provide direct evidence for changes in magnetospheric conditions.

\begin{figure*}
    \centering

    \begin{minipage}{0.48\textwidth}
        \centering
        \textbf{(a) PSR~J1903+1407}\\
        \includegraphics[width=0.95\linewidth]{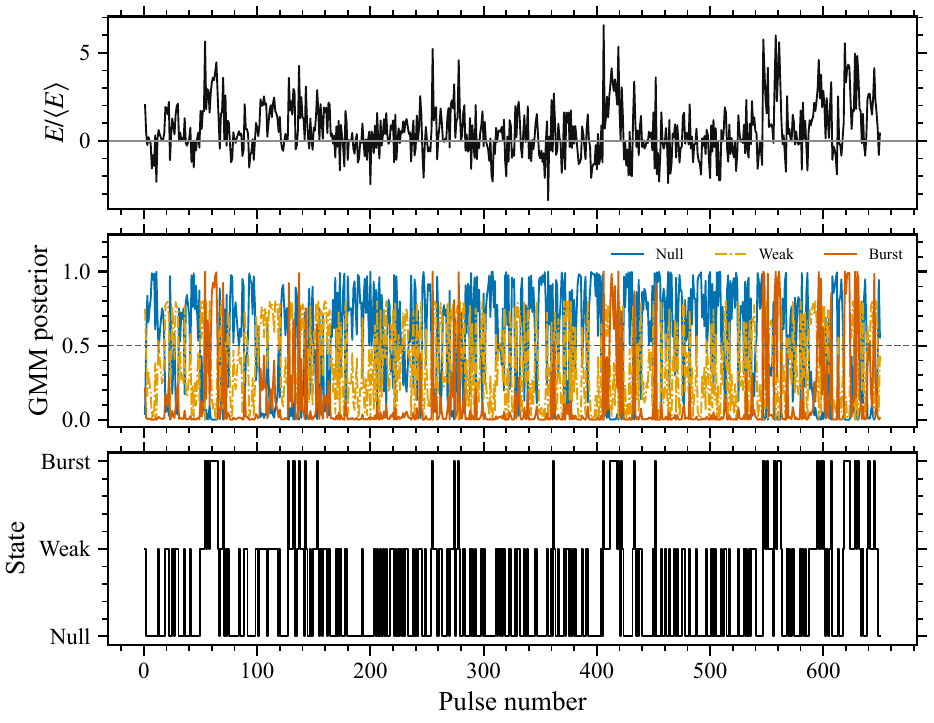}
    \end{minipage}
    \hfill
    \begin{minipage}{0.48\textwidth}
        \centering
        \textbf{(b) PSR~J2131+3642}\\
        \includegraphics[width=0.95\linewidth]{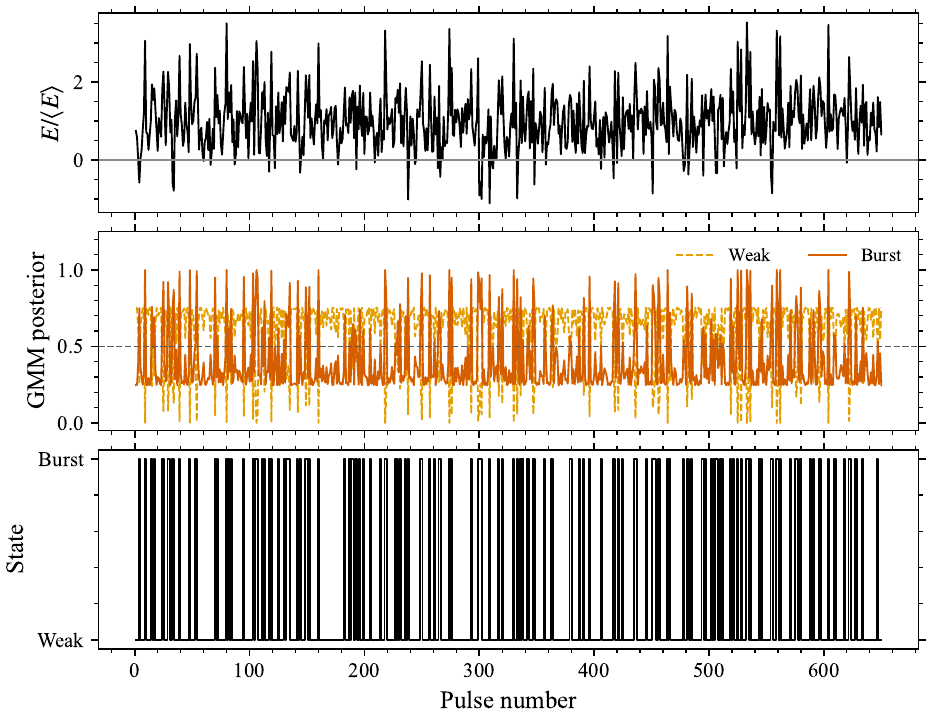}
    \end{minipage}
    \caption{Pulse-by-pulse Gaussian mixture model classification for 
PSR~J1903+1407 and PSR~J2131+3642. For each pulsar, the top panel shows the  normalized pulse energy, $E/\langle E\rangle$, as a function of pulse number. 
The middle panel shows the GMM posterior probability, or responsibility, of each  emission-state component for each pulse. The dashed horizontal line marks a 
posterior probability of 0.5. The bottom panel shows the state  sequence obtained by assigning each pulse to the component with the highest 
posterior probability. PSR~J1903+1407 is classified into null emission, weak, and  burst states, whereas PSR~J2131+3642 is described by weak and burst states.}
    \label{fig:gmm_classification}
\end{figure*}

We analyzed individual pulse sequences for all five long-period sources. Figures~\ref{fig:pulse_stack_timed} and~\ref{fig:pulse_stack_additional} show the pulse stacks and integrated profiles; Figures~\ref{fig:energy_timed} and~\ref{fig:energy_additional} show the energy distributions and GMM decompositions; 
Figures~\ref{fig:gmm_classification} shows the pulse-by-pulse GMM classification; Figures~\ref{fig:average_profiles_timed} and~\ref{fig:average_profiles_additional} show the state-dependent profiles; and Figures~\ref{fig:state_durations_timed} and~\ref{fig:state_durations_additional} show the state-duration distributions. 
All five sources show detectable single-pulse emission, but with markedly different degrees of intermittency, ranging from persistent weak emission to frequent null emission states.

\subsection{Methods for Energy Distribution Analysis}

We determined pulse energy distributions using the method of \citet{Ritchings1976}. For each pulsar, the on-pulse window was defined as the phase range where pulse intensities exceeded $3\sigma_{\rm off}$, with $\sigma_{\rm off}$ as the standard deviation from the off-pulse region. 
This threshold balances sensitivity to weak emission and noise rejection \citep{Kaplan2018}. After baseline subtraction, we summed the on-pulse intensities for each pulse. 
Off-pulse energy was calculated similarly, with an off-pulse window containing the same number of phase bins. 
We normalized pulse energies by the mean on-pulse energy, $\langle E \rangle$, and constructed histograms of the relative pulse energy, $E/\langle E \rangle$.

We characterized emission states using both the classical nulling-fraction method and GMM. In the classical approach of \citet{Ritchings1976}, the off-pulse energy histogram is scaled by a trial nulling fraction and subtracted from the on-pulse histogram. 
The preferred nulling fraction minimizes excess counts at negative pulse energies, where null pulses are expected. While widely used, this method relies on histogram binning and assumes that negative-energy values in the on-pulse distribution arise from null pulses \citep{Kaplan2018}. 
For weak pulsars, radiometer noise can shift intrinsically weak emission pulses below zero, leading to overestimation of the nulling fraction \citep{Ritchings1976, Wang2007, Kaplan2018}.

To quantify the probability that an individual pulse belongs to the null component, we calculated the nulling responsibility for each pulse energy $E_j$,

\begin{equation}
P_{\rm null}(E_j) =
\frac{
p_0 \mathcal{N}(E_j \mid \mu_0, \sigma_0)
}{
\sum_{i} p_i \mathcal{N}(E_j \mid \mu_i, \sigma_i)
}.
\end{equation}
Here, $p_i$ is the mixture weight of the $i$th component, and $\mathcal{N}(E_j \mid \mu_i, \sigma_i)$ is the Gaussian probability density with mean $\mu_i$ and standard deviation $\sigma_i$.
The index $j$ includes all fitted pulse-energy components, such as null, weak-emission, and burst components when present. This probabilistic classification follows \citet{Kaplan2018} and \citet{Anumarlapudi2023},
allowing null pulses to be identified without relying solely on a fixed energy threshold.

Figures~\ref{fig:energy_timed} and~\ref{fig:energy_additional} present normalized pulse-energy histograms and GMM fits for the five sources.
Black solid histograms represent on-pulse energy distributions, while grey dashed histograms represent off-pulse distributions. Green dashed curves display the fitted off-pulse noise distributions.
Blue dash-dotted curves represent null components, yellow dashed curves represent weak-emission components, and orange solid curves represent burst components. The solid black curves show the total fitted model. 
Posterior distributions for representative three-state GMM fits for PSR~J0000+6252 and PSR~J1903+1407 are provided in Appendix~\ref{app:corner_plots} and shown in Figure~\ref{fig:corner_plots}.

\begin{table*}
\centering
\setlength{\tabcolsep}{10pt}
\renewcommand{\arraystretch}{1.25}

\caption{
Dedicated long-duration observations used for the single-pulse analysis. $T_{\rm obs}$ is the on-source duration and $N_{\rm pulse}$ is the
number of observed rotations. RFI-contaminated rotations were excluded before calculating pulse energies, $NF_{\rm classic}$, and the GMM
parameters. The quantities $NF_{\rm classic}$, $f_{\rm null}$, $f_{\rm weak}$, and $f_{\rm burst}$ are computed from the RFI-flagged pulse selection, not directly from $N_{\rm pulse}$. 
The GMM weights sum to unity, and their uncertainties are $1\sigma$ bootstrap standard deviations. No distinct GMM null component is identified for PSR~J2131+3642. These data are independent of the monthly 240-s timing scans described in Section~\ref{sec:obs}.
}
\label{tab:emission_state_summary}

\begin{tabular}{lcccccccc}
\hline\hline
Pulsar
& $P$
& $T_{\rm obs}$
& $N_{\rm pulse}$
& Epoch
& $NF_{\rm classic}$
& $f_{\rm null}$
& $f_{\rm weak}$
& $f_{\rm burst}$ \\
J name & (s) & (s) & & (MJD)
& (\%) & (\%) & (\%) & (\%) \\
\hline

J0000+6252
& 1.552 & 2400 & 1547 & 60784
& $60.86 \pm 2.07$
& $66.0 \pm 1.5$
& $25.1 \pm 1.6$
& $8.9 \pm 0.8$ \\

J1903+1407
& 1.285 & 2400 & 1868 & 60783
& $58.56 \pm 1.79$
& $43.2 \pm 4.0$
& $37.4 \pm 3.2$
& $19.3 \pm 2.0$ \\

J2131+3642
& 1.113 & 2220 & 1996 & 60782
& $27.93 \pm 1.20$
& \textemdash
& $67.0 \pm 7.1$
& $33.0 \pm 7.1$ \\

J1502+4653
& 1.750 & 3600 & 2057 & 60555
& $51.75 \pm 1.61$
& $53.3 \pm 1.2$
& $28.6 \pm 3.2$
& $18.1 \pm 3.2$ \\

J2112+4058
& 4.061 & 4800 & 1182 & 60615
& $35.94 \pm 1.79$
& $38.2 \pm 1.5$
& $49.3 \pm 2.4$
& $12.4 \pm 2.2$ \\

\hline
\hline
\end{tabular}
\end{table*}

\subsection{Results for Individual Pulsars}\label{subsec:individual_pulsars}

This section summarizes the normalized energy distributions, GMM classifications, state-dependent profiles, and state-duration statistics for the five-source sample. 
Pulses were assigned to the GMM component with the highest posterior probability, and state durations were measured as contiguous sequences of pulses assigned to the same state. 
Emission-state fractions are listed in Table~\ref{tab:emission_state_summary}, with GMM parameters and duration statistics in Appendix~\ref{app:gmm_run_lengths} and Table~\ref{tab:state_run_lengths}.

\subsubsection{PSR~J0000+6252}

As shown in Figure~\ref{fig:pulse_stack_timed}(a), PSR~J0000+6252 displays large pulse-to-pulse intensity changes within a narrow pulse-phase range. 
The integrated profile is asymmetric, with a dominant main peak and weaker trailing emission.
The normalized on-pulse energy distribution in Figure~\ref{fig:energy_timed}(a) is concentrated near zero with a positive-energy tail. The exploratory three-component GMM has mixture weights $f_{\rm null}=66.0\pm1.5\%$, $f_{\rm weak}=25.1\pm1.6\%$, and $f_{\rm burst}=8.9\pm0.8\%$, whereas the classical method gives $NF_{\rm classic}=60.86\pm2.07\%$. Because BIC favors two components for this source, the weak--bright subdivision is model-dependent; the robust result is the prominence of low-energy pulses.

\begin{figure*}
    \centering
    \includegraphics[width=0.95\textwidth]{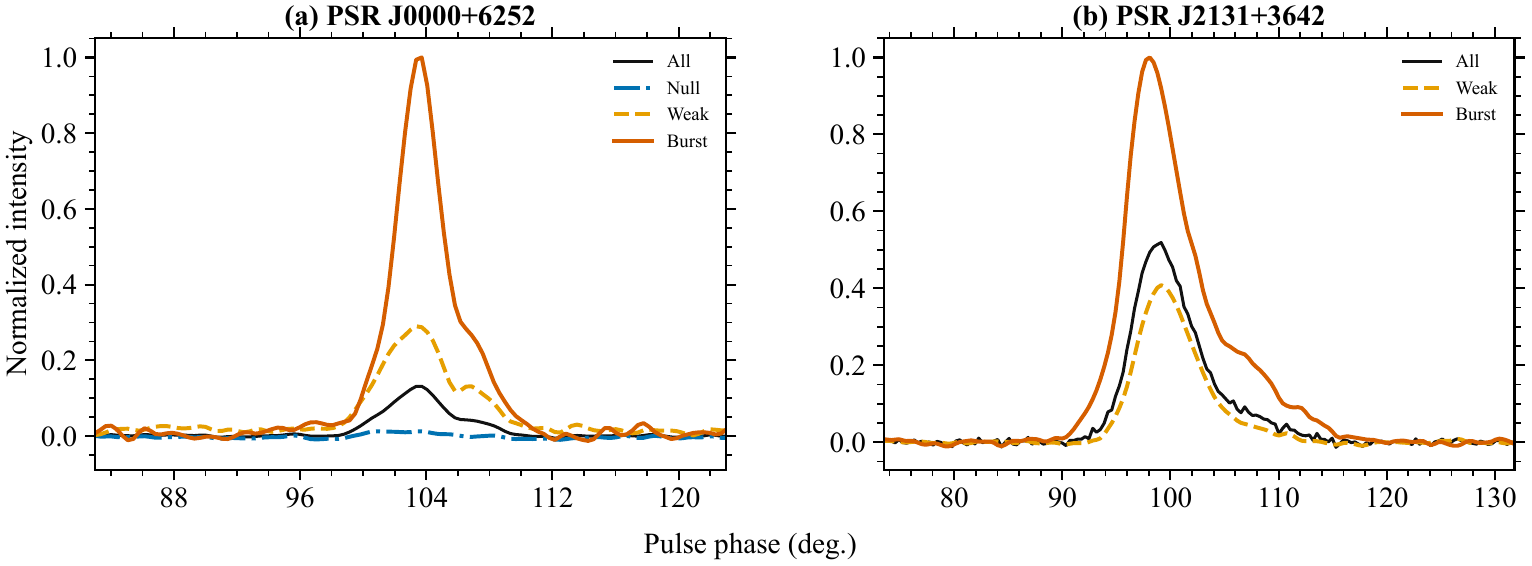}
    \caption{State-dependent average pulse profiles for (a) PSR~J0000+6252 and (b) PSR~J2131+3642. Black curves show the profiles formed from all pulses. PSR~J0000+6252 has null, weak, and burst emission states, whereas PSR~J2131+3642 has only weak and burst emission states. Each panel is normalized by its burst-state peak.}
    \label{fig:average_profiles_timed}
\end{figure*}

\begin{figure*}
    \centering
    \includegraphics[width=0.95\textwidth]{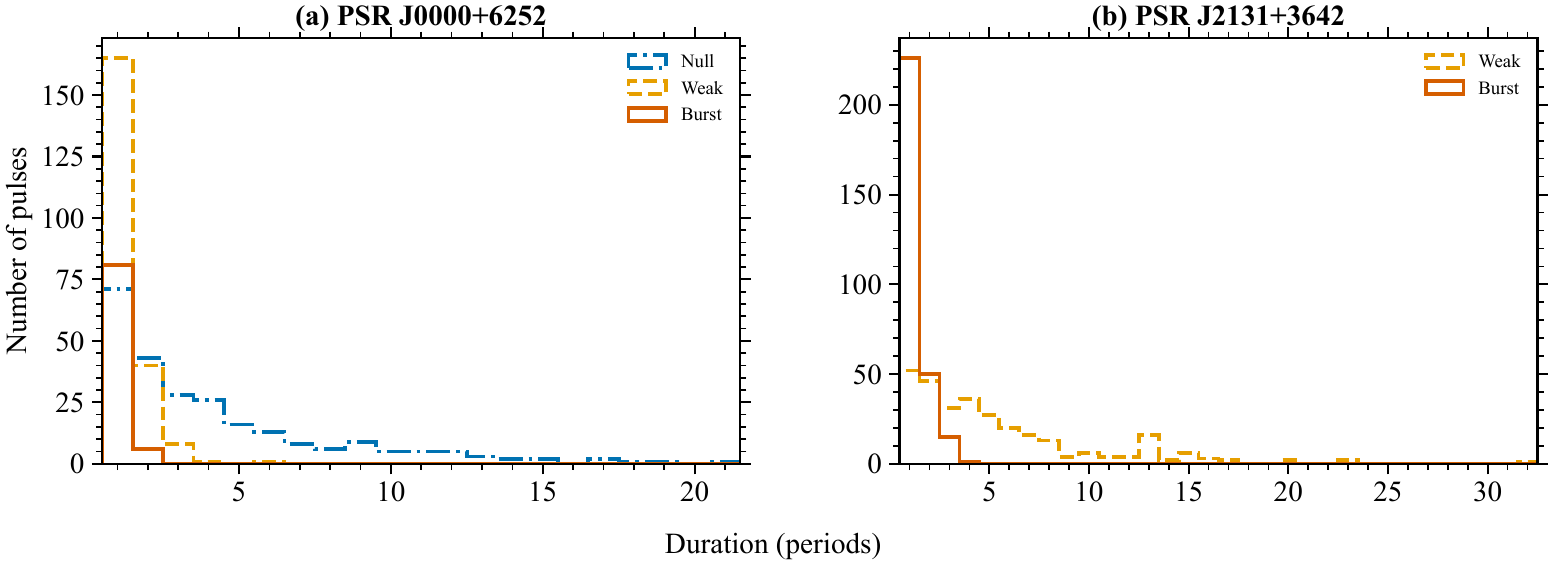}
    \caption{Emission-state duration histograms for (a) PSR~J0000+6252 and (b) PSR~J2131+3642. Durations are measured as the number of consecutive rotations assigned to each GMM state. The line colors and styles follow the state-profile figures.}
    \label{fig:state_durations_timed}
\end{figure*}

The state-dependent profiles in Figure~\ref{fig:average_profiles_timed}(a) reinforce this picture: the burst and weak profiles align with the integrated profile in pulse phase but differ strongly in amplitude, while the null-state profile remains close to the off-pulse baseline. Thus, the GMM states primarily trace changes in emission strength rather than pulse longitude. 
The duration histogram in Figure~\ref{fig:state_durations_timed}(a) shows 1042 null-state pulses grouped into 247 contiguous blocks, with a maximum block length of 21 rotations. The weak and burst selections contain 279 and 93 pulses grouped into 215 and 87 blocks, respectively; their mean block lengths are 1.30 and 1.07 pulses and their maximum block lengths are 6 and 2 pulses.

PSR~J0000+6252 is therefore dominated by null-state emission: the null component accounts for about two thirds of the observation and forms blocks as long as 21 rotations, whereas weak and burst blocks have mean lengths of 1.30 and 1.07 rotations. 
This behavior resembles that of strong nulling pulsars such as PSR~B0835$-$41 and PSR~B1112+50 \citep{Gajjar2012}, and is consistent with models in which state switching reflects changes in magnetospheric particle acceleration or plasma supply \citep{Kerr2014, Young2015}.

\begin{figure*}
    \centering
    \includegraphics[width=0.95\textwidth]{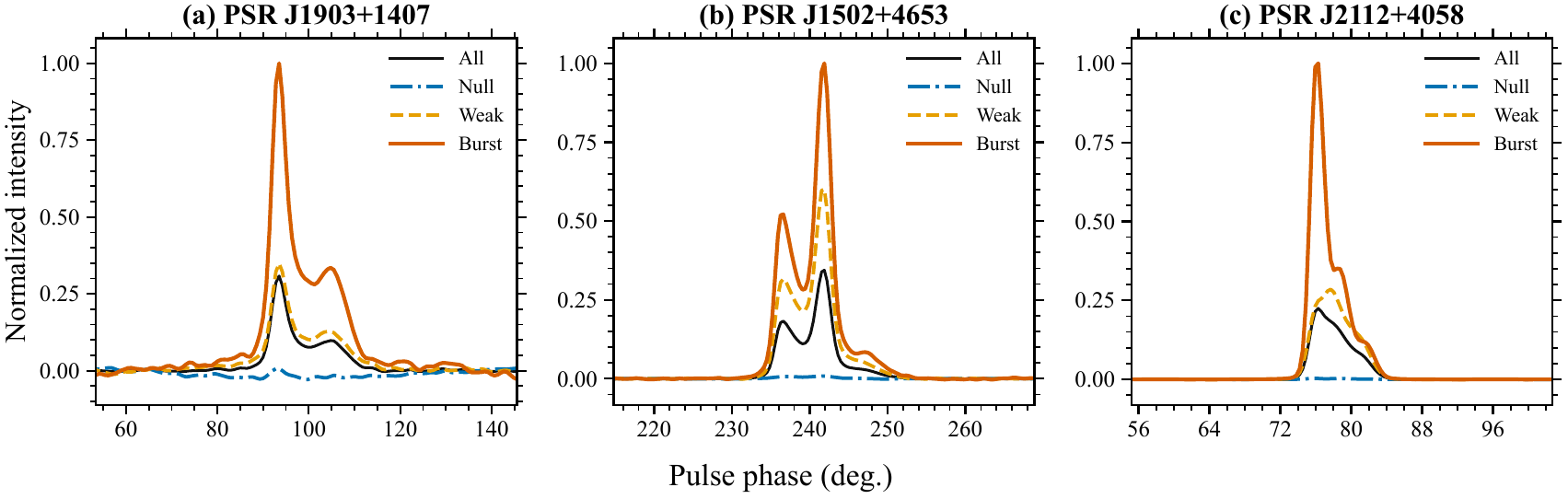}
    \caption{Similar to Figure~\ref{fig:average_profiles_timed}, but for (a) PSR~J1903+1407, (b) PSR~J1502+4653, and (c) PSR~J2112+4058. Each panel is normalized by its burst-state peak.} 
    \label{fig:average_profiles_additional}
\end{figure*}

\subsubsection{PSR~J1903+1407}

The pulse stack for PSR~J1903+1407 in Figure~\ref{fig:pulse_stack_additional}(a) displays strong pulse-to-pulse variability and clear intermittent behavior. Many rotations show weak or null emission pulses, with occasional bright pulses spanning a wide range of energies. The integrated profile features a narrow leading component and weaker trailing emission.

The on-pulse energy distribution in Figure~\ref{fig:energy_additional}(a) is broad, extending from negative to high positive energies. The exploratory three-component mixture weights are $f_{\rm null}=43.2\pm4.0\%$, $f_{\rm weak}=37.4\pm3.2\%$, and $f_{\rm burst}=19.3\pm2.0\%$, compared with $NF_{\rm classic}=58.56\pm1.79\%$. BIC favors two components and the lowest-energy selection retains significant emission, so it must not be interpreted as a proven null population. The larger classical value is consistent with weak emission overlapping the noise distribution \citep{Kaplan2018}.

Figure~\ref{fig:gmm_classification}(a) shows frequent pulse-to-pulse transitions among the null emission, weak, and burst states. PSR~J1903+1407 rarely remains in one state for many rotations. In Figure~\ref{fig:average_profiles_additional}(a), the burst-state profile is much stronger than the weak-state profile and contains both a prominent leading peak and enhanced trailing emission. The weak-state profile has a similar shape but lower amplitude, while the null-state profile is assessed quantitatively against the off-pulse energy distribution above. The phase alignment of the weak and burst profiles indicates that state changes mainly modulate emission strength; the enhanced trailing component during burst states may trace an additional profile component.

\begin{figure*}
    \centering
    \includegraphics[width=0.95\textwidth]{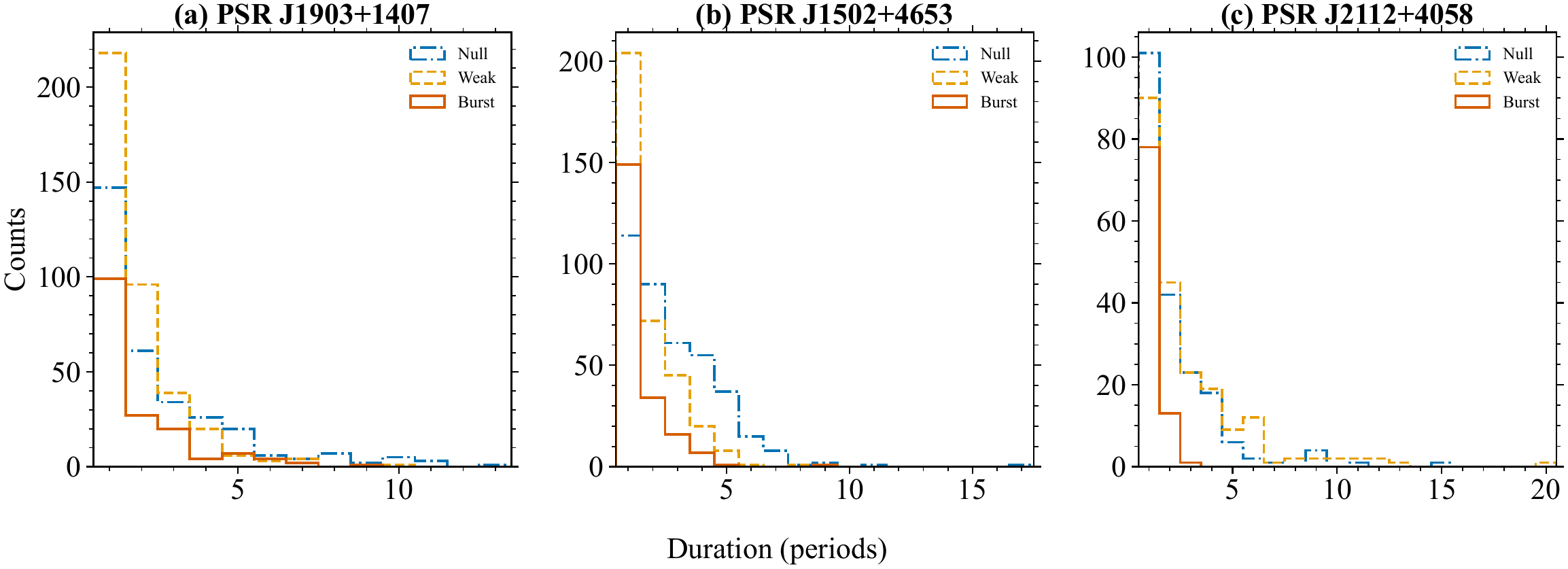}
    \caption{Similar to Figure~\ref{fig:state_durations_timed}, but for (a) PSR~J1903+1407, (b) PSR~J1502+4653, and (c) PSR~J2112+4058.}
    \label{fig:state_durations_additional}
\end{figure*}

The state-duration distribution in Figure~\ref{fig:state_durations_additional}(a) quantifies the switching timescale. The null selection contains 809 pulses grouped into 316 contiguous blocks, with a mean block length of 2.56 pulses and a maximum of 13. 
The weak and burst selections contain 702 and 311 pulses grouped into 388 and 164 blocks, respectively; their mean block lengths are 1.81 and 1.90 pulses and their maxima are 10 and 9 pulses. Compared with PSR~J0000+6252, PSR~J1903+1407 has shorter null intervals and more frequent transitions among states.

\subsubsection{PSR~J2131+3642}

As shown in Figure~\ref{fig:pulse_stack_timed}(b), PSR~J2131+3642 shows strong pulse-to-pulse modulation but more persistent emission than PSR~J0000+6252 and PSR~J1903+1407. Emission is concentrated near the main-pulse phase and detected across many rotations, and the integrated profile is broad and asymmetric.

The on-pulse energy distribution in Figure~\ref{fig:energy_timed}(b) is described by two GMM components, which we label weak emission and burst emission. The GMM does not identify a distinct null component; its fitted mixture fractions are $f_{\rm weak}=67.0\pm7.1\%$ and $f_{\rm burst}=33.0\pm7.1\%$. The weak component has $\mu=0.786$, $\sigma=0.533$, and weight 0.670, while the burst component has $\mu=1.508$, $\sigma=0.866$, and weight 0.330.
The classical method nonetheless returns a nonzero nulling fraction, $NF_{\rm classic}=27.93\pm1.20\%$ (Table~\ref{tab:emission_state_summary}); in the absence of a separate low-energy GMM component for this source, we attribute this to the lowest-energy weak-state pulses scattering below the classical on/off-pulse threshold rather than to a physically distinct null population.

The pulse-by-pulse classification in Figure~\ref{fig:gmm_classification}(b) illustrates this behavior directly. Most pulses are assigned to the weak emission state, while the burst posterior becomes dominant during high-energy excursions.

The state-dependent profiles in Figure~\ref{fig:average_profiles_timed}(b) show weak-state and burst emission at the same pulse phase as the integrated profile. The burst profile has a higher peak amplitude and a more pronounced trailing shoulder, while the weak-state profile retains the same overall shape at lower amplitude.

The duration histogram in Figure~\ref{fig:state_durations_timed}(b) shows that the weak emission state is more persistent than the burst state. The weak state has a mean duration of 5.34 pulses, a median of 4.0 pulses, and a maximum of 32 pulses. In contrast, the burst state has a mean duration of 1.28 pulses, a median of 1.0 pulse, and a maximum of 4 pulses.

Overall, the five sources span a broad range of single-pulse behavior. PSR~J2131+3642 is described by weak and burst emission states, with no distinct null component, whereas PSRs~J0000+6252, J1903+1407, J1502+4653, and J2112+4058 include a distinct null component in the adopted GMM classification. The largest null-state fractions occur in PSR~J0000+6252 and PSR~J1502+4653, while PSR~J2112+4058 shows a lower but still significant low-energy component. This diversity illustrates the range of magnetospheric conditions that can sustain coherent radio emission among long-period sources \citep{Sheikh2021}.

\begin{figure*}[ht]
    \centering
    \includegraphics[width=0.95\textwidth]{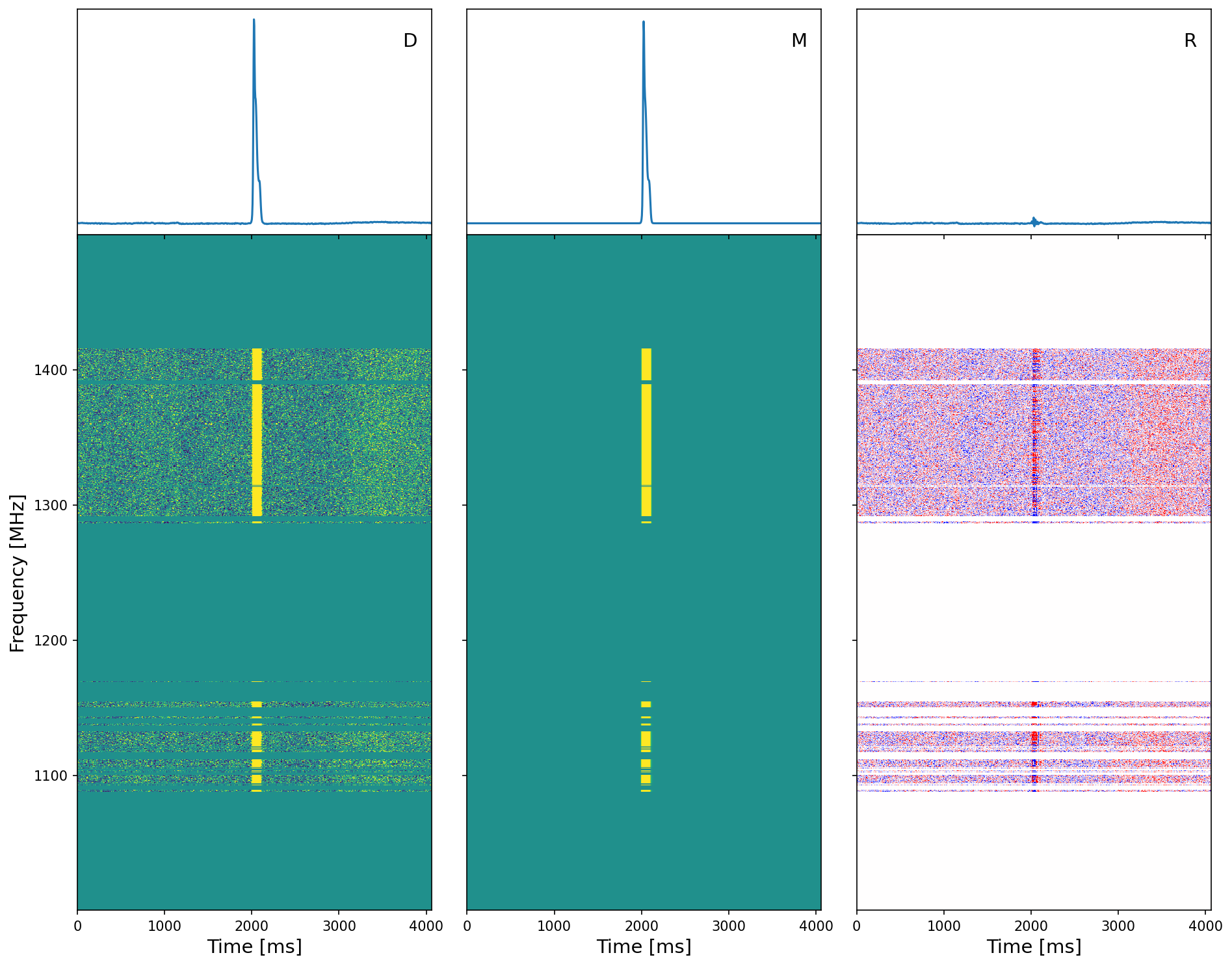}
    \caption{Dynamic spectrum of PSR~J2112+4058 and the best-fitting scattering model. 
    The left panel shows the observed dynamic spectrum, the middle panel shows the best-fitting model from \texttt{Fitburst}\citep{Fonseca2024},  and the right panel shows the residuals after subtracting the model from the data.} 
    \label{fig:j2112_fitburst}
\end{figure*}  

\subsubsection{PSR~J1502+4653}

Figure~\ref{fig:pulse_stack_additional}(b) shows that PSR~J1502+4653 changes frequently among its three emission states: the GMM null fraction is $53.3\pm1.2\%$, and the mean null-, weak-, and burst-block lengths are 2.81, 1.76, and 1.48 rotations, respectively. 
Compared with PSR~J2112+4058, it spends a larger fraction of the observation in the null emission state. 
The state-dependent profiles remain confined to the main pulse window, indicating intensity changes without measurable phase shifts.

The on-pulse energy distribution in Figure~\ref{fig:energy_additional}(b) is strongly non-Gaussian and requires multiple components. 
The GMM separates the distribution into null emission, weak-emission, and burst-emission components. The classical method yields $NF_{\rm classic}=51.75\pm1.61\%$, 
closely matching the GMM null-state fraction of $f_{\rm null}=53.3\pm1.2\%$. The remaining pulses are divided between a weak state ($f_{\rm weak}=28.6\pm3.2\%$) 
and a burst state ($f_{\rm burst}=18.1\pm3.2\%$), showing that detected emission spans both moderate and bright states.

The null emission state is the most common of the three, with weak and burst pulses interspersed.
In Figure~\ref{fig:average_profiles_additional}(b), the burst-state profile has the highest amplitude, the weak-state profile is fainter but covers the same phase range, 
and the null-state profile is assessed quantitatively against the off-pulse energy distribution above. The state-dependent profiles also reveal a multi-component emission structure: 
a weaker leading component precedes the dominant main peak, and a faint trailing shoulder is visible after it. 
These components are strongest in the burst state and weaker in the weak state. The aligned profiles show that pulse phase remains stable while the relative amplitudes of existing profile components vary.

The duration distribution in Figure~\ref{fig:state_durations_additional}(b) confirms that emission states are typically short. 
The null emission state spans 1082 pulse periods, with a mean duration of 2.81 pulses and a maximum of 17. The weak state spans 617 periods, averaging 1.76 pulses and reaching up to 8 pulses. 
The burst state occurs over 307 periods, with a mean of 1.48 pulses and a maximum of 9 pulses. PSR~J1502+4653 is therefore dominated by frequent, short-timescale switching rather than long stable modes.

\subsubsection{PSR~J2112+4058}

The pulse stack of PSR~J2112+4058 in Figure~\ref{fig:pulse_stack_additional}(c) shows alternating null-state rotations, weak pulses, and occasional bright bursts. Most enhanced emission is confined to a narrow pulse-phase range. 
The state-dependent profiles in Figure~\ref{fig:average_profiles_additional}(c) show a compact main component with no distinct secondary component. The state-duration measurements quantify the switching: the mean null-, weak-, and burst-block lengths are 2.20, 2.76, and 1.16 rotations, respectively.

The on-pulse energy distribution in Figure~\ref{fig:energy_additional}(c) is broad and asymmetric, so a single Gaussian is insufficient. A three-component GMM separates the pulses into null emission, weak-emission, and burst-emission components. The classical on-pulse/off-pulse comparison yields $NF_{\rm classic}=35.94\pm1.79\%$, while the GMM weights are $f_{\rm null}=38.2\pm1.5\%$, $f_{\rm weak}=49.3\pm2.4\%$, and $f_{\rm burst}=12.4\pm2.2\%$. The close agreement between the classical and GMM null-state fractions implies that the low-energy component is relatively well separated from active emission, although some overlap with the weakest pulses may remain.

The pulse-by-pulse GMM classification reveals frequent transitions among the three emission states. The weak-emission state is the dominant active state, whereas burst pulses are less common and typically short-lived. In the state-dependent profiles, the weak and burst profiles peak at nearly the same pulse phase and differ mainly in amplitude; the null-state profile is assessed quantitatively against the off-pulse energy distribution above. State changes therefore mainly affect intensity rather than pulse location.

The duration histogram in Figure~\ref{fig:state_durations_additional}(c) shows short-lived states. The null emission state spans 437 pulse periods, with a mean duration of 2.20 pulses and a maximum of 15. The weak state spans 582 periods, averaging 2.76 pulses and reaching up to 20 pulses. The burst state occurs in 107 periods, usually isolated, with a mean of 1.16 pulses and a maximum of 3. PSR~J2112+4058 therefore exhibits rapid emission-state switching rather than long-lived mode changes.

PSR~J2112+4058 also exhibits significant multipath scattering. We fitted the frequency-dependent pulse broadening using \texttt{Fitburst}
\citep{Fonseca2024}. Adopting the four-component model, we obtain a scattering timescale of
$\tau_{\rm sc}=5.84\pm0.18$~ms at the reference frequency $\nu_{\rm ref}=1.25$~GHz. The scattering index was fixed at
$\alpha_{\rm sc}=-4.4$, while the fitted dispersion measure was
$\mathrm{DM}=124.46\pm0.10$~pc\,cm$^{-3}$. The observed dynamic spectrum, best-fitting model, and residuals are shown in Figure~\ref{fig:j2112_fitburst}.

The fitted scattering broadens the intrinsically narrow pulse and redistributes some of its intensity across the pulse window, particularly at the lower-frequency end of the observing band. This broadening should be considered when interpreting the detailed shapes of the state-dependent total-intensity profiles.

\begin{figure*}
    \centering
    \includegraphics[width=0.95\textwidth]{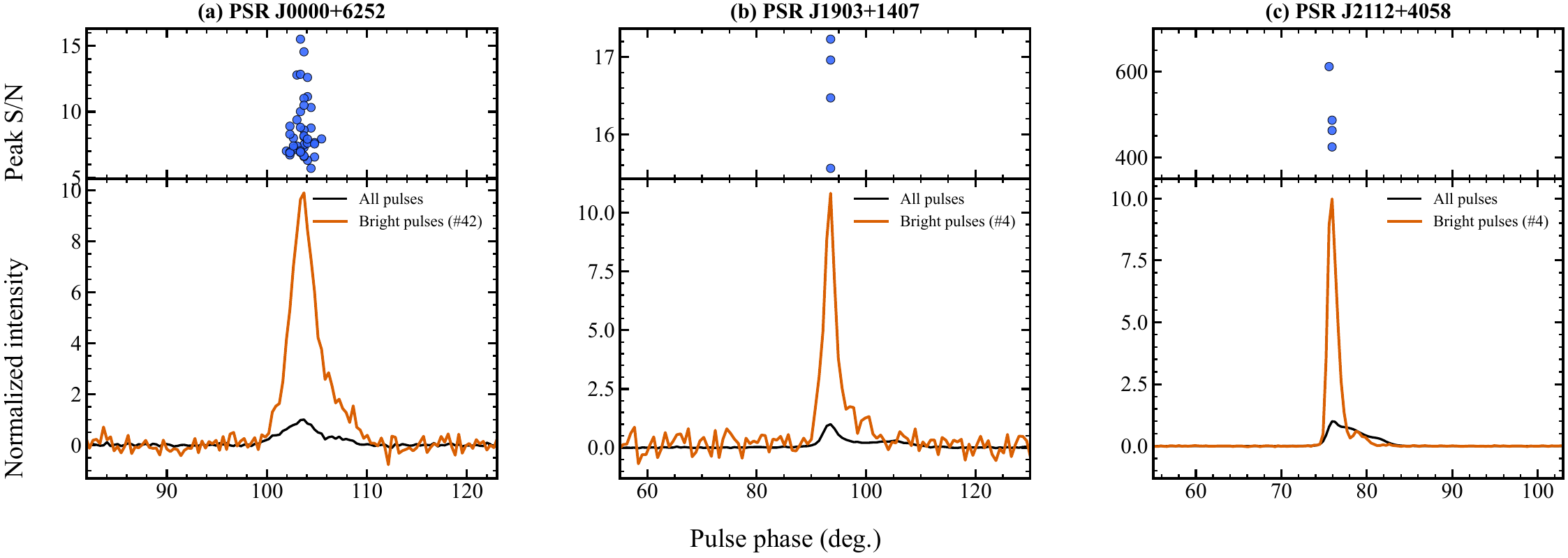}
    \caption{Pulse-phase distribution of the bright pulses from PSRs~J0000+6252, J1903+1407, and J2112+4058. 
    In each panel, the upper sub-panel shows the peak signal-to-noise ratio of individual bright pulses as a function of pulse phase, while the lower sub-panel compares the normalized average profile of all pulses with that of the bright pulses.}
    \label{fig:peak_snr_bright_profiles}
\end{figure*}

\subsection{Bright-pulse detection}\label{subsec:bright_pulses}

A search for bright single pulses was conducted to investigate sporadic increases in radio emission and their potential connection to magnetospheric variability. Following \citet{Tedila2022}, bright pulses were defined as those with peak intensity at least 10 times that of the integrated average pulse profile.

\begin{equation}
I_{\rm peak}^{\rm SP} \geq 10\,I_{\rm peak}^{\rm avg},
\end{equation}
where $I_{\rm peak}^{\rm SP}$ is the peak intensity of an individual single pulse, and $I_{\rm peak}^{\rm avg}$ is the peak intensity of the integrated average profile. 
This operational threshold selects the high-intensity tail of ordinary single-pulse emission, including events often called bright or giant-micropulse-like pulses; it does not by itself establish that an event is a classical giant pulse \citep[e.g.][]{Weltevrede2006,Tedila2022,Tedila2025}.

For each detected bright pulse, the peak signal-to-noise ratio was measured as

\begin{equation}
({\rm S/N})_{\rm peak}
=
\frac{I_{\rm peak}^{\rm SP}-\mu_{\rm off}}{\sigma_{\rm off}},
\end{equation}
where $\mu_{\rm off}$ and $\sigma_{\rm off}$ are the mean and root mean square of the off-pulse region. The corresponding peak flux density was estimated using the radiometer equation.

\begin{equation}
S_{\rm peak}
=
\frac{\beta T_{\rm sys}}
{G\sqrt{n_{\rm pol}\,\Delta\nu\,t_{\rm bin}}}
({\rm S/N})_{\rm peak},
\end{equation}
where $\beta$ is the digitization loss factor, $T_{\rm sys}$ is the system temperature, $G$ is the telescope gain, $n_{\rm pol}$ is the number of summed polarizations, $\Delta\nu$ is the observing bandwidth, and $t_{\rm bin}$ is the phase-bin width. 
This calculation provides the maximum instantaneous brightness of each pulse, rather than the pulse-averaged flux density. 
The waiting time between successive bright pulses was also analyzed. For two consecutive bright pulses with pulse numbers $N_i$ and $N_{i+1}$, the waiting time is

\begin{equation}
\Delta N_i = N_{i+1} - N_i .
\end{equation}

Alternatively, $\Delta t_i = P\,\Delta N_i$, where $P$ is the pulsar spin period. The waiting-time distribution assesses whether bright pulses occur randomly or show recurrence. 
42 bright pulses were detected in PSR~J0000+6252, and four bright pulses were detected in each of PSRs~J1903+1407 and J2112+4058. No pulses above the bright-pulse threshold were found in PSRs~J2131+3642 or J1502+4653. 
The properties of the detected bright pulses are listed in Appendix~\ref{app:bright_pulses} and Tables~\ref{table:burst1903}, \ref{table:burst2112}, and~\ref{table:burst0001} for PSRs~J1903+1407, J2112+4058, and J0000+6252, respectively.
The bright pulses in PSRs~J0000+6252, J1903+1407, and J2112+4058 are concentrated within the main emission window, as shown in Figure~\ref{fig:peak_snr_bright_profiles}. 
The bright pulses do not display a periodic pattern, indicating they occur sporadically.

\begin{figure*}  
    \centering
    \begin{minipage}{0.48\textwidth}
        \centering
        \includegraphics[width=0.95\linewidth]{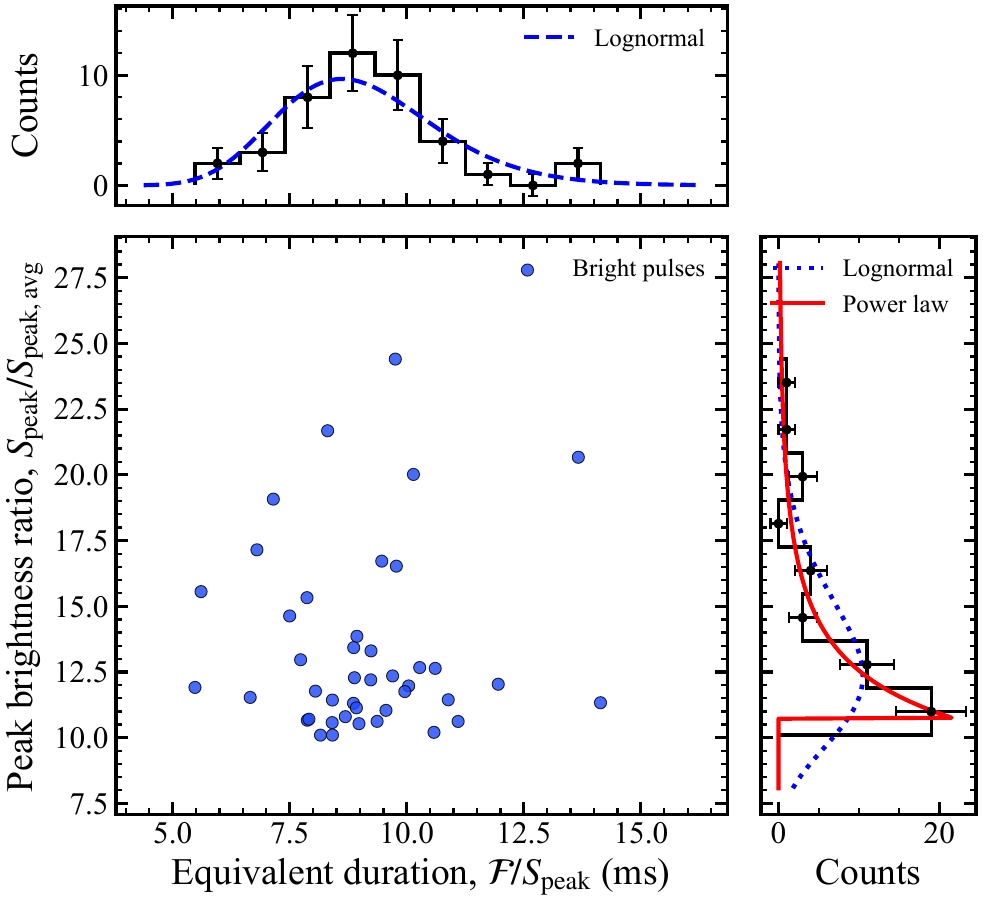}
    \end{minipage}
    \hfill
    \begin{minipage}{0.48\textwidth}
        \centering
        \includegraphics[width=0.95\linewidth]{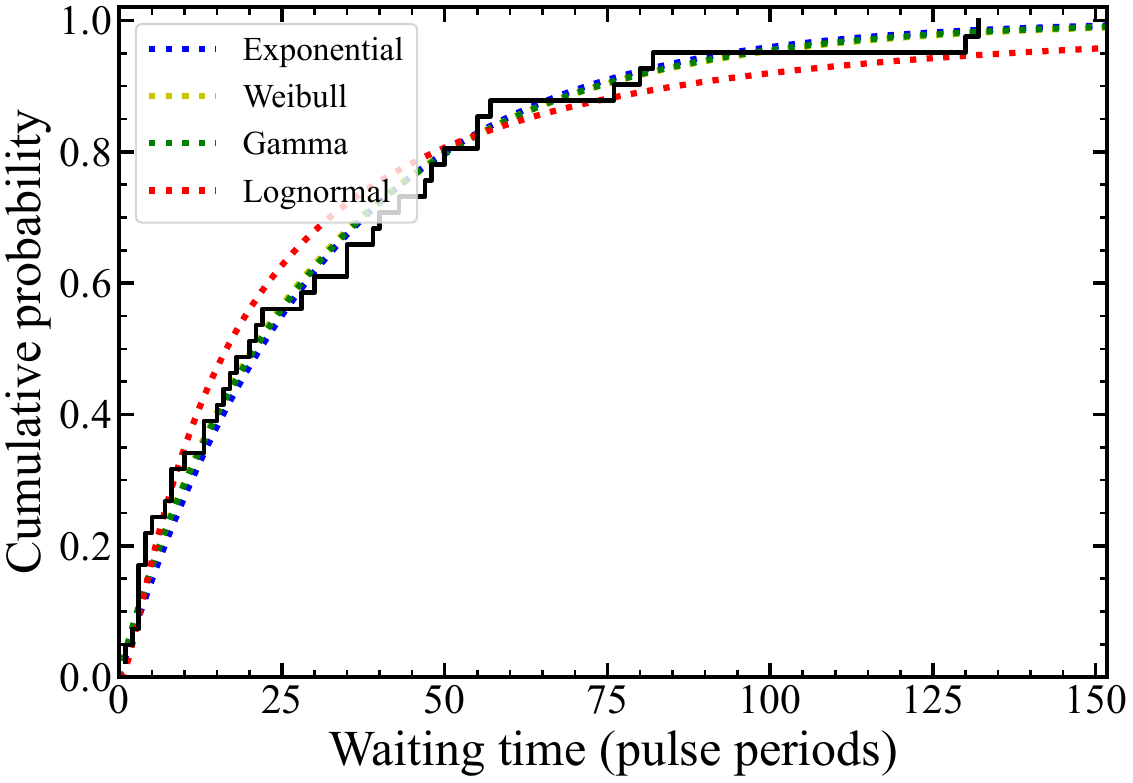}
    \end{minipage}
    \caption{Bright-pulse properties of PSR~J0000+6252 using the 42 pulses selected by the $I_{\rm peak}^{\rm SP} \geq 10I_{\rm peak}^{\rm avg}$ criterion. Panel (a) shows the peak-brightness ratio, $S_{\rm peak}/S_{\rm peak,avg}$, as a function of equivalent duration, $\mathcal{F}/S_{\rm peak}$; its marginal histograms include maximum-likelihood lognormal fits and a maximum-likelihood power-law fit above the stated lower cutoff. Panel (b) shows the empirical cumulative waiting-time distribution between consecutive bright pulses, expressed in units of pulse periods, compared with exponential, Weibull, gamma, and lognormal models.}
    \label{fig:j0001_bright_pulse_properties}
\end{figure*}

Figure~\ref{fig:peak_snr_bright_profiles} shows the pulse-phase distribution of the bright pulses from PSRs~J0000+6252, J1903+1407, and J2112+4058: the upper sub-panel gives the peak signal-to-noise ratio of individual bright pulses versus pulse phase, and the lower sub-panel compares the normalized average profile of all pulses with that of the bright pulses alone. In all three sources the bright pulses are confined to the main emission window, indicating that they arise from the same emission region as the normal pulses rather than from a distinct component. Comparable high-intensity single-pulse events have been reported in PSR~B0656+14 \citep{Weltevrede2006}, PSR~B0950+08 \citep{Singal2012}, and PSR~B0031$-$07 \citep{Tsai2016}, and in classical giant-pulse emitters such as PSR~B1937+21 \citep{McKee2019}.

Physically, such events could represent either the bright tail of the normal single-pulse energy distribution or a distinct giant-pulse-like population, the latter typically marked by narrow widths, restricted pulse longitude, and non-lognormal energy statistics. Because the bright pulses reported here remain within the main emission window, they are more consistent with short-lived enhancements of the normal emission beam than with a separate emission component, possibly reflecting intermittent changes in the radio-emission process or localized changes in the magnetospheric plasma \citep[e.g.][]{Petrova2006}. The larger number of events detected in PSR~J0000+6252 (42, versus four each in PSRs~J1903+1407 and J2112+4058) suggests a higher intrinsic rate of high-intensity pulses, though the measured rate may also depend on sensitivity, observing duration, and the underlying pulse-energy statistics.

The four bright pulses detected in each of PSRs~J1903+1407 and J2112+4058 are too few to permit a reliable waiting-time distribution fit; we therefore restrict such fits to PSR~J0000+6252, and any conclusions for the other two sources remain limited by small-number statistics.

Figure~\ref{fig:j0001_bright_pulse_properties}(a) shows the peak-brightness ratio, $R_{\rm peak}\equiv S_{\rm peak}/S_{\rm peak,avg}$, versus the equivalent duration, $W_{\rm eq}\equiv\mathcal{F}/S_{\rm peak}$, for the 42 bright pulses from PSR~J0000+6252; the marginal distributions shown were fit by maximum likelihood directly to these 42 measurements, and the histograms are for visualization only. The observed sample medians are $R_{\rm peak}=12.00$ and $W_{\rm eq}=8.93$~ms. Two-parameter lognormal fits give $\sigma_W=0.190$ and fitted median $s_W=8.931$~ms for $W_{\rm eq}$ ($D_{\rm KS}=0.086$, $p_{\rm KS}=0.888$), consistent with the observed median, and $\sigma_R=0.226$ and fitted median $s_R=12.994$ for $R_{\rm peak}$ ($D_{\rm KS}=0.187$, $p_{\rm KS}=0.091$), somewhat above the observed median and indicating weaker agreement with a lognormal model. The upper tail of $R_{\rm peak}$ was additionally fit with a power law,
\begin{equation}
p(R_{\rm peak})=
\frac{\alpha-1}{R_{\min}}
\left(\frac{R_{\rm peak}}{R_{\min}}\right)^{-\alpha},
\qquad R_{\rm peak}\geq R_{\min},
\end{equation}
using the 20th-percentile cutoff $R_{\min}=10.725$, which selects 33 of the 42 events and yields $\alpha=4.982$ ($D_{\rm KS}=0.100$, $p_{\rm KS}=0.867$), indicating that more extreme pulses become rapidly less common.

Figure~\ref{fig:j0001_bright_pulse_properties}(b) shows the waiting-time distribution between successive bright pulses, comprising 41 intervals measured from the 42 detected pulses, with mean $31.20P$ and median $20P$, where $P$ is the pulsar spin period. Compared against exponential, Weibull, gamma, and lognormal models, the empirical cumulative distribution is broadly consistent with stochastic recurrence and does not indicate strictly periodic behavior: the Weibull and gamma distributions give the highest Kolmogorov--Smirnov probabilities ($p\simeq0.93$), and the exponential model performs best by the Akaike information criterion. Given the modest sample size, we interpret these model comparisons descriptively rather than as definitive evidence for the underlying waiting-time process.

\begin{figure*}
    \centering
    \includegraphics[width=2\columnwidth]{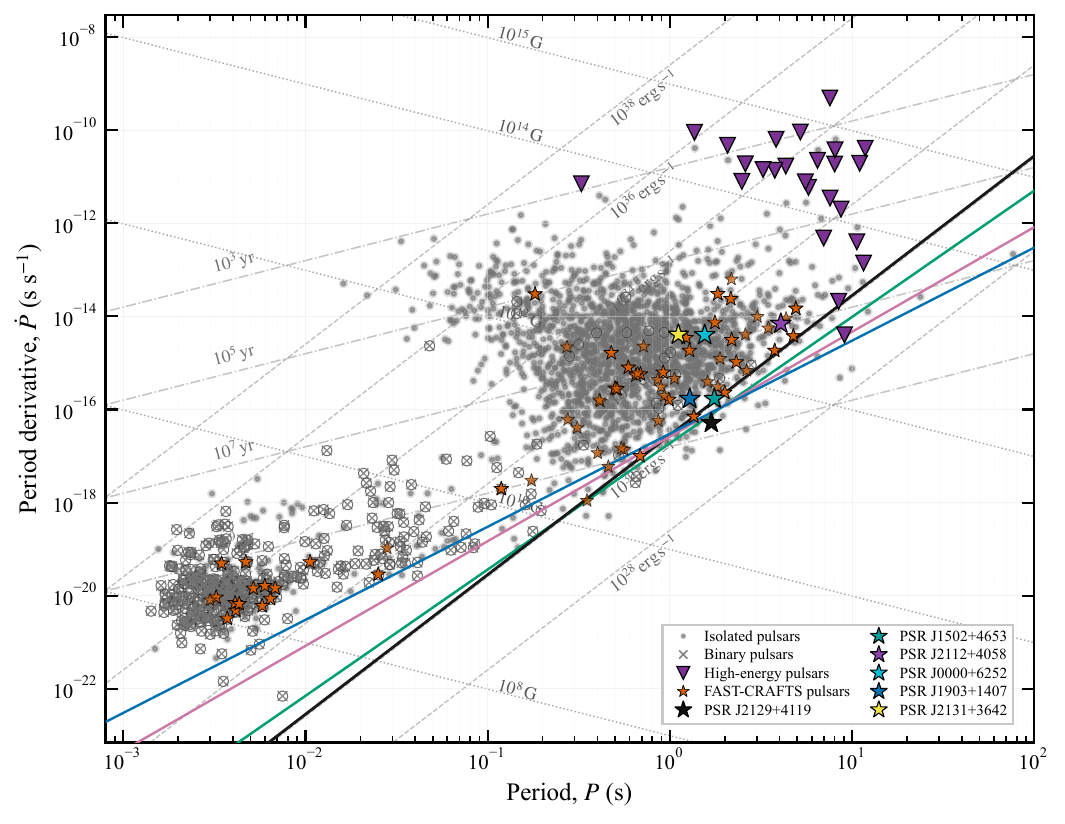}
\caption{
$P$--$\dot{P}$ diagram of the five long-period pulsars studied here, shown relative to the ATNF pulsar population \citep{Manchester2005}. 
Gray circles, gray crosses, purple inverted triangles, and orange stars denote isolated, binary, high-energy, and FAST-CRAFTS pulsars, respectively. 
The five target pulsars and the comparison source PSR~J2129+4119 are highlighted with colored stars. Lines of constant characteristic age, surface magnetic field, and spin-down luminosity are shown, 
together with the traditional death line \citep{Ruderman1975}, death valley \citep{Chen1993}, and curvature-radiation limits for vacuum-gap and space-charge-limited-flow models \citep{Zhang2000}.
}
    \label{fig:P_Pdot}
\end{figure*}

\section{Discussion}\label{sec:discussion}

We analyzed phase-connected timing solutions and polarization properties for two long-period FAST-CRAFTS pulsars, PSRs~J0000+6252 and J2131+3642, and extended the single-pulse emission-state analysis to a five-source sample that also includes PSRs~J1903+1407, J1502+4653, and J2112+4058.
The five sources display diverse single-pulse behavior, ranging from steady weak emission to strong nulling and occasional bright pulses.
This diversity makes long-period pulsars useful probes of magnetospheric stability, plasma supply, and emission-state transitions \citep{Rankin1986, Rankin1993, Lyne2010, Perera2015, Basu2018, Wright2022}.

\subsection{Long-period FAST pulsars in the $P$--$\dot{P}$ diagram} \label{subsec:ppdot_discussion}

The two pulsars with phase-connected timing solutions occupy the long-period region of the $P$--$\dot{P}$ diagram, with spin periods of 1.11--1.55~s and period derivatives of $(3.99$--$4.06)\times10^{-15}$~s~s$^{-1}$. 
PSRs~J1903+1407, J1502+4653, and J2112+4058, included here only in the single-pulse analysis, extend the contextual comparison across spin periods of 1.29, 1.75, and 4.06~s, respectively. These parameters classify the sample as slowly rotating, non-recycled radio pulsars. As shown in Figure~\ref{fig:P_Pdot}, they lie among normal isolated pulsars, but near the region where radio emission becomes sensitive to polar-cap potential, 
pair creation, and magnetospheric plasma supply \citep{Ruderman1975, Chen1993, Zhang2000, Lorimer2005, Manchester2005}. 
This sample is well-suited to testing whether coherent radio emission persists in slowly rotating neutron stars.

PSR~J0000+6252 has a period of 1.55~s, a characteristic age of 6.16~Myr, and the lower spin-down luminosity of the two timed sources ($\dot{E}=4.21\times10^{31}$~erg~s$^{-1}$). PSR~J2131+3642 has the shorter period ($P=1.11$~s), a characteristic age of 4.34~Myr, and a spin-down luminosity of $1.16\times10^{32}$~erg~s$^{-1}$.

These differences place the timed pulsars at distinct evolutionary stages, while PSRs~J1502+4653 and J2112+4058 broaden the single-pulse comparison toward more intermittent and longer-period behavior.
Previous studies have shown that nulling is more common among long-period and older pulsars, but with substantial scatter \citep{Ritchings1976, Wang2007, Gajjar2012, BasuMitra2018, Sheikh2021}. 
Our five-source results show substantial diversity in nulling and emission-state behavior, indicating that spin period alone does not determine the presence or strength of nulling.

This conclusion is consistent with recent FAST studies of other long-period CRAFTS pulsars. 
\citet{Tedila2025} found that PSRs~J1945+1211, J2323+1214, and J1900$-$0134, with periods of 1.83--4.75~s, show nulling fractions of about 28--53\%, 
but with different transition patterns, including quasiperiodic nulls, abrupt null--burst switching, gradual transitions, dwarf pulses, bright pulses, and microstructure. 
Similarly, \citet{Tedila2026} showed that PSR~J2129+4119, despite lying below the traditional radio-death line and having a very low spin-down luminosity (the black comparison star in Figure~\ref{fig:P_Pdot}), remains radio active with a low nulling fraction of only $8.13\pm0.51\%$. 
Taken together, these studies show that long-period pulsars near the death-line region follow multiple emission pathways: some become strong nullers, while others remain mostly active but exhibit weak emission, short nulls, drifting, amplitude modulation, or occasional bright pulses.

The $P$--$\dot{P}$ positions provide useful evolutionary context, but single-pulse behavior shows that emission stability depends on several additional factors.
Rotational energy loss, surface magnetic field, pair-production efficiency, magnetospheric current structure, and viewing geometry all likely contribute \citep{Rankin1993, Lyne2010, Timokhin2010, Young2015, Wright2022}. 
In this view, radio death is not a sharp boundary in the $P$--$\dot{P}$ diagram, but a broad transition region shaped by magnetic geometry as well as magnetospheric state \citep{Chen1993, Zhang2000, Szary2014, BasuMitra2018}. 

\subsection{Emission-state fractions and their relation to spin-down parameters}\label{subsec:nulling_parameter_correlation}

Our results for the five sources divide the sample into two observational groups under the Gaussian mixture analysis.
PSRs~J0000+6252, J1903+1407, J1502+4653, and J2112+4058 require three descriptive components: null emission, weak emission, and burst emission. 
PSR~J2131+3642 is described by two states: weak and burst, with no distinct null component. 
This distinction is more informative than a single nulling fraction because weak emission can overlap with the off-pulse noise distribution and may be misclassified as null emission by classical methods \citep{Ritchings1976, Wang2007, Kaplan2018, Anumarlapudi2023}.
Growing evidence across the pulsar population supports this multi-state view: individual sources can combine several coexisting emission phenomena rather than a simple binary on/off (null-versus-burst) dichotomy. \citet{Tedila2026} found that PSR~J2129+4119 exhibits nulls, weak pulses, regular emission, and occasional bright pulses simultaneously, while \citet{Wang2026} identified three coexisting variability phenomena in PSR~J0846$-$3533 -- periodic amplitude modulation, mode switching between two distinct profile states, and intermittent nulling. These results support a multi-component, rather than strictly binary, description of pulsar emission states, consistent with the three- and two-state classifications adopted here.

The source-by-source measurements are given once in Section~\ref{subsec:individual_pulsars} and Table~\ref{tab:emission_state_summary}; here we emphasize only the comparison. PSR~J0000+6252 has the largest null-state fraction and the longest null emission runs, whereas PSR~J2131+3642 is described by weak and burst emission states. PSRs~J1903+1407, J1502+4653, and J2112+4058 occupy intermediate regimes, with frequent short transitions rather than long stable modes. The disagreement between the classical and mixture estimates is largest for PSR~J1903+1407, consistent with its significant residual signal in the lowest-energy selection.

Within the two timed sources, PSR~J0000+6252 has the lower spin-down luminosity and stronger null-emission behavior, while PSR~J2131+3642 has the higher spin-down luminosity and strong amplitude modulation without a distinct null component. This two-object contrast is suggestive but is insufficient to establish a population correlation between $\dot E$ and nulling. Similar behavior in intermittent pulsars is generally interpreted in terms of changes in plasma supply or magnetospheric current configuration \citep{Gajjar2012, Kramer2006, Lyne2010, Timokhin2010, Young2015}.

The comparison with PSR~J2129+4119 further strengthens this interpretation. 
That source lies below the traditional death line and has $\dot{E}=0.65\times10^{30}$~erg~s$^{-1}$, yet it continues to be dominated by active emission and has a low nulling fraction of $8.13\pm0.51\%$ \citep{Tedila2026}. 
It also shows null, weak, regular, and bright pulses, demonstrating that a pulsar can operate close to the pair-production threshold without entering a predominantly null emission state. 
Similarly, the dwarf pulses detected in PSR~J2323+1214 suggest that some apparently null-like intervals may contain weak residual emission rather than complete radio silence \citep{Tedila2025}. 
This supports multi-component energy-distribution modeling: weak emission is physically meaningful and should not be automatically grouped with nulls. This approach improves the interpretation of emission states across long-period pulsars.

Overall, comparison among the five sources yields three main conclusions.
First, nulling does not increase monotonically with spin period: the longest-period source, PSR~J2112+4058, shows only a moderate low-energy fraction. 
Second, the sources with distinct low-energy components span a range of periods and emission strengths, from the large null-state fraction of PSR~J0000+6252 to the shorter null intervals of PSRs~J1502+4653 and J2112+4058. 
Third, the lower-$\dot{E}$ timed source, PSR~J0000+6252, shows strong nulling, whereas PSR~J2131+3642 has a higher $\dot{E}$ and shows strong modulation with a two-state weak--burst classification and no distinct null component. 
Emission-state classification is likely determined by a combination of period, characteristic age, magnetic field, spin-down power, pair-production efficiency, magnetospheric state, and viewing geometry.

\subsection{Geometry, lack of inclination angle constraints, and interpretation of state switching}\label{subsec:geometry_state_switching}

The role of emission geometry, particularly the inclination angle ($\alpha$) and impact angle ($\beta$), cannot be robustly quantified for the two-pulsar polarization sample. PSR~J0000+6252 has too few usable PA points, while PSR~J2131+3642 has a monotonic segment that can be fitted formally but covers too little pulse longitude to break the strong RVM parameter degeneracies.
In the rotating vector model, the magnetic inclination angle $\alpha$ and impact angle $\beta$ can be constrained from the shape and slope of the PA curve \citep{Radhakrishnan69, Komesaroff1970, EverettWeisberg2001}. 
Accordingly, the PSR~J2131+3642 RVM parameters are reported as a diagnostic fit but are not used in the subsequent correlations or physical interpretation.
Similar limitations are common in long-period pulsars with narrow profiles, low linear-polarization signal-to-noise ratios, or complex orthogonal polarization modes \citep{Rankin1993, Manchester2005, Johnston2005, Noutsos2009, EverettWeisberg2001}. 
Since we lack reliable estimates of inclination and impact angles, we do not attempt to correlate these geometric parameters with nulling behavior.

The geometric interpretation is instead based on profile morphology and phase alignment. 
In all five sources, the weak- and burst-state profiles peak at approximately the same pulse longitude as the integrated profile. 
The identified states therefore appear to be primarily intensity variations within the same emission window rather than separate beams at different rotational phases.
In PSRs~J1903+1407 and J2131+3642, the burst-state profile shows stronger trailing emission than the weak-state profile. 
Burst states may therefore enhance an existing profile component or activate a weak secondary component rather than produce emission at a new longitude.

This interpretation is consistent with mode-changing and intermittent pulsars, in which changes in pulse intensity and profile shape are linked to variations in the global magnetospheric state \citep{Backer1970, Rankin1986, Kramer2006, Lyne2010, Lower2025}. 
In these models, nulls and mode changes are not independent phenomena; both can result from changes in plasma generation, polar-cap acceleration, or magnetospheric current closure \citep{Timokhin2010, Young2015, Melrose2017, Wright2022}. 
The present data support this view because the emission states vary mainly in intensity and duration, and the emission longitude remains stable.

The comparison with other FAST long-period pulsars further supports a component-dependent interpretation. 
In PSR~J2323+1214, the nulling fraction differs between the leading and trailing components, and dwarf pulses are detected only in the trailing component \citep{Tedila2025}. 
In PSR~J2129+4119, emission after nulls shows enhanced trailing-component intensity, suggesting gradual magnetospheric reactivation rather than a purely geometric origin \citep{Tedila2026}. 
These results are relevant to PSRs~J1903+1407 and J2131+3642, in which the burst-state profiles also exhibit enhanced trailing emission. 
State switching may involve selectively strengthening or weakening specific emission components within the same radio beam.

The detection of bright pulses further supports this interpretation. 
Bright pulses are observed in PSRs~J0000+6252, J1903+1407, and J2112+4058, all of which show a distinct low-energy component in the GMM analysis. 
These pulses occur within the main emission window and are best described as high-intensity pulses from the normal emission beam, rather than classical giant pulses from a separate high-energy magnetospheric region. 
Classical giant pulses are typically associated with very high magnetic fields at the light cylinder and narrow phase windows aligned with high-energy emission \citep[e.g., PSR~J1824--2452A;][]{Knight2006}.
The bright pulses reported here are more consistent with sporadic strengthening of normal radio emission, as observed in other long-period pulsars \citep{Weltevrede2006, Gajjar2012, Tedila2022, Tedila2025}. 
Bright pulses also occur in long-period sources with different nulling fractions and should be interpreted as brief enhancements of ordinary radio emission, not as a feature unique to strong nullers.

\section{Summary}\label{sec:summary}

We report timing and polarization measurements for two long-period FAST-CRAFTS pulsars, along with single-pulse analyses for a sample of five sources.
The main results are summarized as follows:

\begin{itemize}

\item We obtained phase-connected timing solutions for PSRs~J0000+6252 and J2131+3642 over baselines of 511 and 414 days, respectively. Their timing properties confirm that they are isolated, non-recycled pulsars.

\item The two pulsars show different polarization properties. PSR~J0000+6252 has too few reliable position-angle measurements to fit a rotating-vector model. 

\item PSR~J2131+3642 shows a short monotonic position-angle segment, but its limited longitude coverage leaves the magnetic inclination and impact angles poorly constrained.

\item In the adopted descriptive GMM classification, PSRs~J0000+6252, J1903+1407, J1502+4653, and J2112+4058 show null, weak, and burst components. PSR~J2131+3642 is described by weak and burst states, with no distinct null component.

\item The comparison between classical nulling fractions and GMM classifications shows that weak intrinsic emission can bias classical nulling estimates. 

\item PSR~J0000+6252 is the most strongly nulling source in the sample, with a GMM null-state fraction of $66.0\pm1.5\%$ and null-state durations extending up to 21 rotations.

\item The weak- and burst-state profiles generally peak at similar pulse longitudes. The state changes therefore primarily modulate the intensity and relative strength of existing profile components within the same observed emission window.

\item We detected 42 bright pulses in PSR~J0000+6252 and four bright pulses in each of PSRs~J1903+1407 and J2112+4058. These events occur within the normal emission window and are consistent with sporadic enhancements of the ordinary radio-emission beam.

\item For PSR~J2112+4058, we measured a scattering timescale of $\tau_{\rm sc}=5.84\pm0.18$~ms at a reference frequency of 1.25~GHz.

\end{itemize}

\section*{Acknowledgments}

We thank the FAST team and the CRAFTS survey collaboration for providing the data used in this study. 
The observations were conducted under projects PT2024--0036 and PT2020--0220. This work made use of the Five-hundred-meter Aperture Spherical radio Telescope (FAST), 
a Chinese national mega-science facility operated by the National Astronomical Observatories, Chinese Academy of Sciences.

R.Y. is supported by the National Natural Science Foundation of China (NSFC) project No.~12288102, the National Key Program for Science and Technology Research and Development No.~2022YFC2205201, the National SKA Program of China No.~2020SKA0120200, 
the National Natural Science Foundation of China (NSFC) projects Nos.~12041303 and 12041304, the Major Science and Technology Program of Xinjiang Uygur Autonomous Region Nos.~2022A03013-2 and 2022A03013-4, and NSFC Grant No.~12573051. 
This research is partly supported by the Operation, Maintenance and Upgrading Fund for Astronomical Telescopes and Facility Instruments, budgeted by the Ministry of Finance of China (MOF) and administered by the CAS.

\software{DSPSR \citep{Straten11}, PSRCHIVE \citep{Hotan04}, PRESTO \citep{Ransom2011}, TEMPO \citep{Nice2015}, TEMPO2 \citep{Hobbs06}, Fitburst \citep{Fonseca2024}}



\clearpage
\onecolumngrid
\appendix
\setcounter{table}{0}
\renewcommand{\thetable}{\Alph{section}\arabic{table}}

\section{Bright Pulse Information in Long-Period Pulsars}
\label{app:bright_pulses}

A bright pulse is defined as an individual pulse whose baseline-subtracted peak intensity is at least ten times the peak intensity of the average profile formed
from all pulses. This appendix lists the bright pulses identified in PSR~J1903+1407, PSR~J2112+4058, and PSR~J0000+6252. The individual bright-pulse measurements are
given in Tables~\ref{table:burst1903}, \ref{table:burst2112}, and \ref{table:burst0001}, respectively. For each pulse, we provide the pulse number, peak signal-to-noise
ratio (S/N), arrival time in Modified Julian Date (MJD), width at 10\% of the peak intensity ($W_{10}$), peak flux density, and fluence.

These pulsars were observed at a central frequency of 1.25~GHz. Because we report the maximum instantaneous brightness of each bright pulse, the listed flux
density is the peak single-bin flux density computed from the peak S/N using the radiometer equation. Quoted uncertainties are $1\sigma$ statistical errors.

For completeness, the mean flux density of a periodic pulse profile can be estimated as
\begin{equation}
S_{\rm mean}
=
\frac{\beta\,({\rm S/N})\,T_{\rm sys}}
{G\sqrt{n_{\rm pol}\,b\,t_{\rm obs}}}
\left(\frac{W_{10}}{P-W_{10}}\right)^{1/2},
\end{equation}
where $\beta$ is the system loss factor, $T_{\rm sys}$ is the system temperature, $G$ is the telescope gain, $n_{\rm pol}$ is the number of summed
polarizations, $b$ is the observing bandwidth, $t_{\rm obs}$ is the integration time, $P$ is the pulse period, and $W_{10}$ is the pulse width at 10\% of the
peak. This expression estimates a pulse-averaged flux density. In the tables below, however, we quote the maximum instantaneous brightness, computed as
\begin{equation}
S_{\rm peak}
=
\frac{\beta\,({\rm Peak\ S/N})\,T_{\rm sys}}
{G\sqrt{n_{\rm pol}\,B\,\Delta t_{\rm bin}}},
\end{equation}
where $\Delta t_{\rm bin}$ is the phase-bin width.

\setlength{\tabcolsep}{6.5pt}

\begin{longtable}{ccccccc}
\caption{Bright pulse peak-brightness information for PSR~J1903+1407.}\label{table:burst1903} \\
\toprule
\textbf{S. No.} &
\textbf{Pulse No.} &
\textbf{Peak S/N} &
\textbf{Pulse Arrival Time} &
\textbf{$W_{10}$} &
\textbf{Peak Flux Density} &
\textbf{Fluence} \\
&&& \textbf{(MJD)} & \textbf{(ms)} & \textbf{(mJy)} & \textbf{(Jy\,ms)} \\
\midrule
\endfirsthead

\caption[]{Bright pulse peak-brightness information for PSR~J1903+1407 -- continued.} \\
\toprule
\textbf{S. No.} &
\textbf{Pulse No.} &
\textbf{Peak S/N} &
\textbf{Pulse Arrival Time} &
\textbf{$W_{10}$} &
\textbf{Peak Flux Density} &
\textbf{Fluence} \\
&&& \textbf{(MJD)} & \textbf{(ms)} & \textbf{(mJy)} & \textbf{(Jy\,ms)} \\
\midrule
\endhead

\midrule
\multicolumn{7}{r}{\textit{Continued on next page}} \\
\endfoot

\bottomrule
\endlastfoot

1 & 721  & 15.56 & $60783.011129889775475$ & $65.26 \pm 5.86$  & $15.64 \pm 1.01$ & $0.162 \pm 0.014$ \\
2 & 993  & 17.23 & $60783.015175353626546$ & $45.18 \pm 10.13$ & $17.32 \pm 1.01$ & $0.242 \pm 0.016$ \\
3 & 1087 & 16.96 & $60783.016573418339249$ & $57.73 \pm 8.52$  & $17.05 \pm 1.01$ & $0.213 \pm 0.014$ \\
4 & 1352 & 16.47 & $60783.020514770985756$ & $32.63 \pm 12.75$ & $16.56 \pm 1.01$ & $0.230 \pm 0.015$ \\
\end{longtable}
\onecolumngrid

\vspace{-0.5em}
{\footnotesize
\noindent\textit{Note.} For PSR~J1903+1407, peak flux densities were computed from the peak single-bin S/N using the radiometer equation, with $\Delta t_{\rm bin}$
equal to the phase-bin width of 2.510~ms. The fluence values are integrated over the on-pulse window.\par}

\setlength{\tabcolsep}{6.5pt}

\begin{longtable}{ccccccc}
\caption{Bright pulse peak-brightness information for PSR~J2112+4058.}\label{table:burst2112} \\
\toprule
\textbf{S. No.} &
\textbf{Pulse No.} &
\textbf{Peak S/N} &
\textbf{Pulse Arrival Time} &
\textbf{$W_{10}$} &
\textbf{Peak Flux Density} &
\textbf{Fluence} \\
&&& \textbf{(MJD)} & \textbf{(ms)} & \textbf{(mJy)} & \textbf{(Jy\,ms)} \\
\midrule
\endfirsthead

\caption[]{Bright pulse peak-brightness information for PSR~J2112+4058 -- continued.} \\
\toprule
\textbf{S. No.} &
\textbf{Pulse No.} &
\textbf{Peak S/N} &
\textbf{Pulse Arrival Time} &
\textbf{$W_{10}$} &
\textbf{Peak Flux Density} &
\textbf{Fluence} \\
&&& \textbf{(MJD)} & \textbf{(ms)} & \textbf{(mJy)} & \textbf{(Jy\,ms)} \\
\midrule
\endhead

\midrule
\multicolumn{7}{r}{\textit{Continued on next page}} \\
\endfoot

\bottomrule
\endlastfoot

1 & 96  & 486.56 & $60615.514843491568172$ & $19.83 \pm 1.14$ & $389.10 \pm 0.80$ & $4.880 \pm 0.019$ \\
2 & 184 & 611.82 & $60615.518979445543664$ & $27.76 \pm 1.14$ & $489.27 \pm 0.80$ & $6.743 \pm 0.019$ \\
3 & 515 & 462.59 & $60615.534536272425612$ & $27.76 \pm 1.14$ & $369.94 \pm 0.80$ & $6.714 \pm 0.019$ \\
4 & 640 & 424.10 & $60615.540411207053694$ & $27.76 \pm 1.14$ & $339.16 \pm 0.80$ & $5.571 \pm 0.019$ \\
\end{longtable}
\onecolumngrid

\vspace{-0.5em}
{\footnotesize
\noindent\textit{Note.} For PSR~J2112+4058, peak flux densities were computed
from the peak single-bin S/N using the radiometer equation, with $\Delta t_{\rm bin}$
equal to the phase-bin width of 3.966~ms.\par}

\setlength{\tabcolsep}{5pt}

\begin{longtable}{ccccccc}
\caption{Bright pulse peak-brightness information for PSR~J0000+6252.}\label{table:burst0001} \\
\toprule
\textbf{S. No.} &
\textbf{Pulse No.} &
\textbf{Peak S/N} &
\textbf{Pulse Arrival Time} &
\textbf{$W_{10}$} &
\textbf{Peak Flux Density} &
\textbf{Fluence} \\
&&& \textbf{(MJD)} & \textbf{(ms)} & \textbf{(mJy)} & \textbf{(Jy\,ms)} \\
\midrule
\endfirsthead

\caption[]{Bright pulse peak-brightness information for PSR~J0000+6252 -- continued.} \\
\toprule
\textbf{S. No.} &
\textbf{Pulse No.} &
\textbf{Peak S/N} &
\textbf{Pulse Arrival Time} &
\textbf{$W_{10}$} &
\textbf{Peak Flux Density} &
\textbf{Fluence} \\
&&& \textbf{(MJD)} & \textbf{(ms)} & \textbf{(mJy)} & \textbf{(Jy\,ms)} \\
\midrule
\endhead

\midrule
\multicolumn{7}{r}{\textit{Continued on next page}} \\
\endfoot

\bottomrule
\endlastfoot

1  & 96   & 8.10  & $60784.087959991884418$ & $22.74 \pm 1.25$ & $10.47 \pm 1.29$ & $0.108 \pm 0.008$ \\
2  & 151  & 6.56  & $60784.088948222444742$ & $21.22 \pm 0.84$ & $8.49 \pm 1.29$  & $0.069 \pm 0.008$ \\
3  & 164  & 8.00  & $60784.089181804214604$ & $22.74 \pm 1.80$ & $10.35 \pm 1.29$ & $0.092 \pm 0.008$ \\
4  & 199  & 6.91  & $60784.089810678211506$ & $21.22 \pm 1.68$ & $8.94 \pm 1.29$  & $0.072 \pm 0.008$ \\
5  & 275  & 12.79 & $60784.091176233167062$ & $18.19 \pm 2.34$ & $16.54 \pm 1.29$ & $0.118 \pm 0.007$ \\
6  & 291  & 8.90  & $60784.091463718425075$ & $21.22 \pm 1.55$ & $11.52 \pm 1.29$ & $0.103 \pm 0.008$ \\
7  & 346  & 7.10  & $60784.092451948992675$ & $22.74 \pm 1.31$ & $9.19 \pm 1.29$  & $0.100 \pm 0.008$ \\
8  & 476  & 9.38  & $60784.094787766684021$ & $13.64 \pm 3.06$ & $12.14 \pm 1.29$ & $0.091 \pm 0.008$ \\
9  & 491  & 7.28  & $60784.095057284110226$ & $18.19 \pm 1.98$ & $9.42 \pm 1.29$  & $0.079 \pm 0.007$ \\
10 & 494  & 7.94  & $60784.095111187598377$ & $13.64 \pm 4.34$ & $10.27 \pm 1.29$ & $0.056 \pm 0.007$ \\
11 & 507  & 12.84 & $60784.095344769368239$ & $22.74 \pm 1.03$ & $16.61 \pm 1.29$ & $0.227 \pm 0.009$ \\
12 & 554  & 7.66  & $60784.096189257303195$ & $19.71 \pm 2.21$ & $9.90 \pm 1.29$  & $0.083 \pm 0.007$ \\
13 & 562  & 8.76  & $60784.096332999928563$ & $19.71 \pm 1.77$ & $11.34 \pm 1.29$ & $0.100 \pm 0.008$ \\
14 & 601  & 10.01 & $60784.097033745238150$ & $22.74 \pm 3.18$ & $12.95 \pm 1.29$ & $0.102 \pm 0.008$ \\
15 & 651  & 6.72  & $60784.097932136661257$ & $22.74 \pm 2.32$ & $8.69 \pm 1.29$  & $0.096 \pm 0.008$ \\
16 & 681  & 10.32 & $60784.098471171513665$ & $22.74 \pm 4.86$ & $13.35 \pm 1.29$ & $0.075 \pm 0.007$ \\
17 & 763  & 11.15 & $60784.099944533445523$ & $13.64 \pm 2.65$ & $14.42 \pm 1.29$ & $0.098 \pm 0.008$ \\
18 & 791  & 6.31  & $60784.100447632641590$ & $22.74 \pm 2.53$ & $8.17 \pm 1.29$  & $0.073 \pm 0.008$ \\
19 & 794  & 11.01 & $60784.100501536129741$ & $22.74 \pm 2.19$ & $14.25 \pm 1.29$ & $0.135 \pm 0.008$ \\
20 & 797  & 8.62  & $60784.100555439610616$ & $22.74 \pm 1.67$ & $11.14 \pm 1.29$ & $0.086 \pm 0.007$ \\
21 & 819  & 7.50  & $60784.100950731837656$ & $21.22 \pm 1.87$ & $9.70 \pm 1.29$  & $0.065 \pm 0.008$ \\
22 & 862  & 8.81  & $60784.101723348459927$ & $22.74 \pm 1.69$ & $11.40 \pm 1.29$ & $0.105 \pm 0.007$ \\
23 & 864  & 7.62  & $60784.101759284116270$ & $16.68 \pm 2.61$ & $9.86 \pm 1.29$  & $0.118 \pm 0.008$ \\
24 & 867  & 7.40  & $60784.101813187604421$ & $22.74 \pm 0.85$ & $9.57 \pm 1.29$  & $0.135 \pm 0.008$ \\
25 & 888  & 6.59  & $60784.102190511999652$ & $21.22 \pm 2.11$ & $8.53 \pm 1.29$  & $0.072 \pm 0.007$ \\
26 & 906  & 8.16  & $60784.102513932914007$ & $18.19 \pm 2.10$ & $10.56 \pm 1.29$ & $0.102 \pm 0.008$ \\
27 & 913  & 10.49 & $60784.102639707707567$ & $21.22 \pm 1.23$ & $13.57 \pm 1.29$ & $0.133 \pm 0.008$ \\
28 & 923  & 7.02  & $60784.102819385996554$ & $21.22 \pm 1.93$ & $9.08 \pm 1.29$  & $0.071 \pm 0.007$ \\
29 & 928  & 5.70  & $60784.102909225133772$ & $18.19 \pm 3.32$ & $7.38 \pm 1.29$  & $0.065 \pm 0.010$ \\
30 & 985  & 8.29  & $60784.103933391357714$ & $22.74 \pm 1.24$ & $10.72 \pm 1.29$ & $0.114 \pm 0.008$ \\
31 & 1002 & 7.22  & $60784.104238844440260$ & $21.22 \pm 1.31$ & $9.34 \pm 1.29$  & $0.089 \pm 0.008$ \\
32 & 1050 & 14.55 & $60784.105101300207025$ & $18.19 \pm 2.01$ & $18.82 \pm 1.29$ & $0.156 \pm 0.008$ \\
33 & 1051 & 7.92  & $60784.105119268031558$ & $19.71 \pm 2.32$ & $10.25 \pm 1.29$ & $0.095 \pm 0.008$ \\
34 & 1183 & 6.65  & $60784.107491021386522$ & $22.74 \pm 1.32$ & $8.60 \pm 1.29$  & $0.091 \pm 0.008$ \\
35 & 1223 & 7.04  & $60784.108209734520642$ & $21.22 \pm 1.53$ & $9.11 \pm 1.29$  & $0.085 \pm 0.007$ \\
36 & 1303 & 7.56  & $60784.109647160796158$ & $21.22 \pm 1.60$ & $9.78 \pm 1.29$  & $0.098 \pm 0.008$ \\
37 & 1307 & 6.94  & $60784.109719032108842$ & $19.71 \pm 2.27$ & $8.97 \pm 1.29$  & $0.078 \pm 0.008$ \\
38 & 1311 & 15.51 & $60784.109790903428802$ & $15.16 \pm 2.90$ & $20.06 \pm 1.29$ & $0.196 \pm 0.008$ \\
39 & 1331 & 6.88  & $60784.110150259992224$ & $19.71 \pm 1.33$ & $8.90 \pm 1.29$  & $0.070 \pm 0.008$ \\
40 & 1339 & 12.61 & $60784.110294002624869$ & $22.74 \pm 2.10$ & $16.31 \pm 1.29$ & $0.165 \pm 0.008$ \\
41 & 1374 & 6.95  & $60784.110922876614495$ & $21.22 \pm 1.82$ & $8.99 \pm 1.29$  & $0.080 \pm 0.008$ \\
42 & 1375 & 7.40  & $60784.110940844446304$ & $19.71 \pm 1.09$ & $9.57 \pm 1.29$  & $0.095 \pm 0.008$ \\
\end{longtable}
\onecolumngrid

\vspace{-0.5em}
{\footnotesize
\noindent\textit{Note.} For PSR~J0000+6252, peak flux densities were computed from the peak single-bin S/N using the radiometer equation, with $\Delta t_{\rm bin}$
equal to the phase-bin width of 1.516~ms. The fluence values are integrated over the on-pulse window.\par}


\section{Posterior Distributions of Emission-State Model Parameters}
\label{app:corner_plots}

\setcounter{figure}{0} 
\renewcommand{\thefigure}{\Alph{section}\arabic{figure}}

\begin{figure}[H]
    \centering

    \begin{minipage}{0.48\textwidth}
        \centering
        \textbf{(a) PSR~J0000+6252}\\[2pt]
        \includegraphics[width=0.95\linewidth]{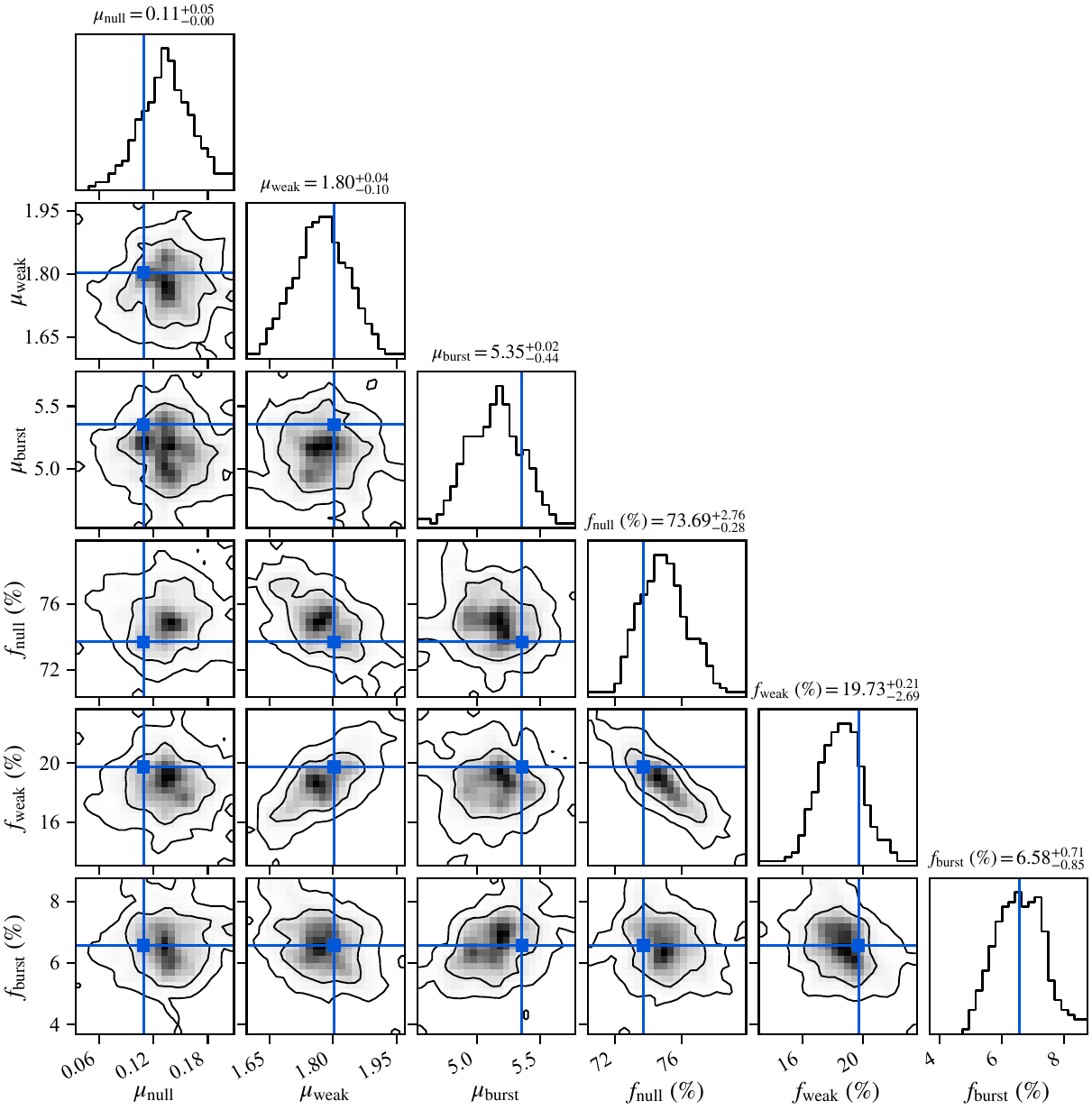}
    \end{minipage}
    \hfill
    \begin{minipage}{0.48\textwidth}
        \centering
        \textbf{(b) PSR~J1903+1407}\\[2pt]
        \includegraphics[width=0.95\linewidth]{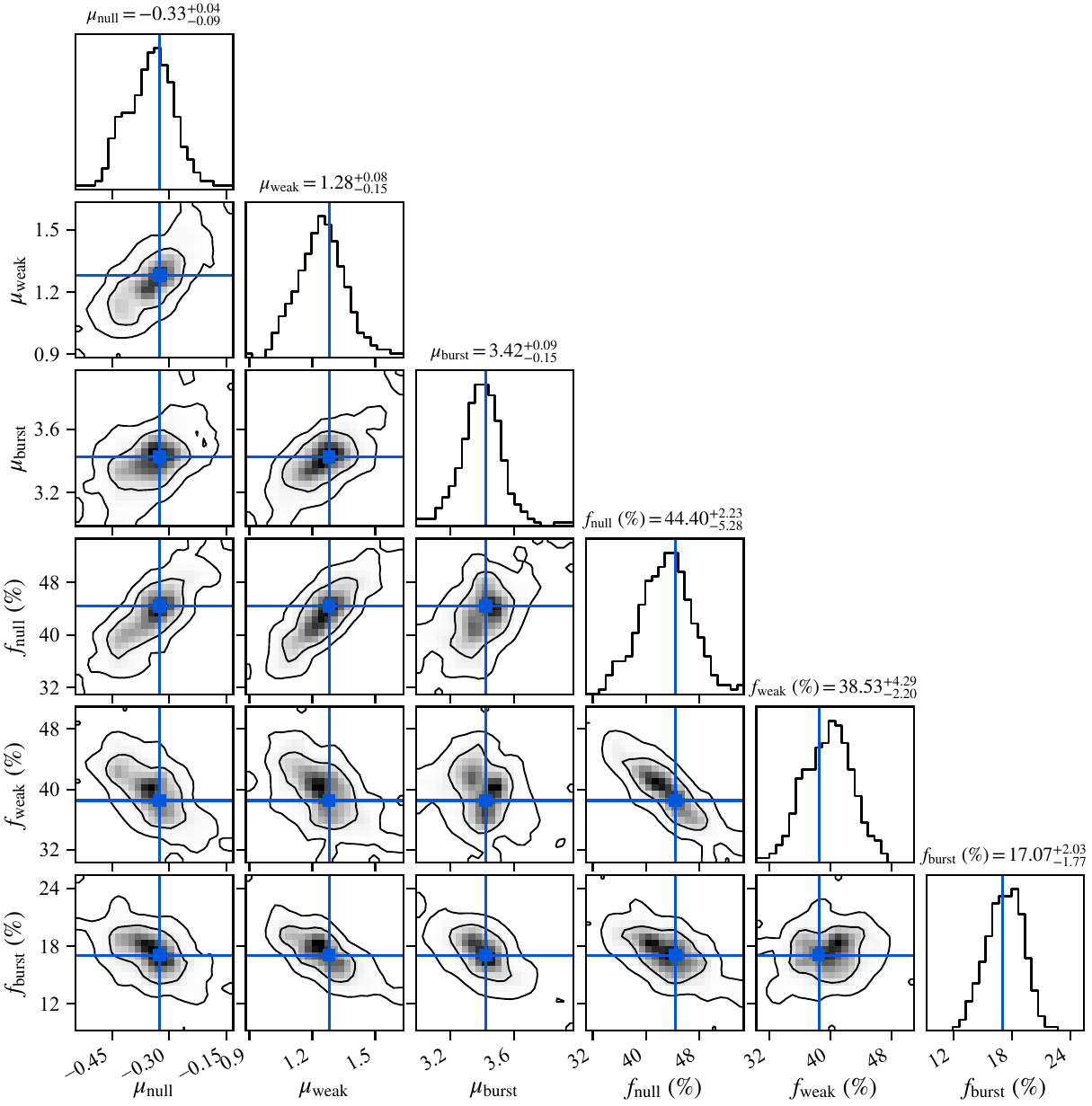}
    \end{minipage}

    \caption{Posterior distributions of the emission-state model parameters
    for PSR~J0000+6252 and PSR~J1903+1407. The diagonal panels show the
    marginalized one-dimensional posterior distributions, while the
    off-diagonal panels show the two-dimensional joint posterior distributions.
    The contours indicate the 1$\sigma$, 2$\sigma$, and 3$\sigma$ credible
    regions.}
    \label{fig:corner_plots}
\end{figure}

Figure~\ref{fig:corner_plots} presents the posterior distributions of the emission-state model parameters for PSR~J0000+6252 and PSR~J1903+1407. The
diagonal panels show the marginalized posterior distribution of each parameter, whereas the off-diagonal panels show the joint posterior distributions for
pairs of parameters. The contours indicate the 1$\sigma$, 2$\sigma$, and 3$\sigma$ credible regions.

\section{State-Duration Properties and Gaussian Mixture Model Parameters}
\label{app:gmm_run_lengths}
\setcounter{table}{0}

Table~\ref{tab:state_run_lengths} summarizes the state-duration properties and Gaussian mixture model parameters of the identified emission states. A state-duration is
defined as a consecutive sequence of pulses assigned to the same state. The parameters $\mu$, $\sigma$, and weight are obtained from the GMM fit to the normalized pulse-energy distribution. 
 
\begin{table}
\centering
\setlength{\tabcolsep}{3pt}
\renewcommand{\arraystretch}{1.25}
\caption{State-duration properties and Gaussian mixture model parameters of the identified emission states. The parameters $\mu$, $\sigma$, and weight are from
the GMM fit to the normalized pulse-energy distribution; weight uncertainties are $1\sigma$ bootstrap standard deviations. RFI-contaminated rotations were excluded before fitting and state assignment. The number of state-durations
refers to the number of consecutive sequences identified for each emission state.}
\label{tab:state_run_lengths}
\begin{tabular}{lcccccccc}
\hline
Pulsar & State & $\mu$ & $\sigma$ & Weight
& Number of & Mean duration & Median duration & Maximum duration \\
       &       &       &          &
& state-durations & (pulses) & (pulses) & (pulses) \\
\hline
J0000+6252 & Null  & 0.109 & 0.878 & $0.660\pm0.015$ & 247 & 4.22 & 3.0 & 21 \\
           & Weak  & 1.803 & 1.181 & $0.251\pm0.016$ & 215 & 1.30 & 1.0 & 6  \\
           & Burst & 5.353 & 3.129 & $0.089\pm0.008$ & 87  & 1.07 & 1.0 & 2  \\
J2131+3642 & Weak  & 0.786 & 0.533 & $0.670\pm0.071$ & 293 & 5.34 & 4.0 & 32 \\
           & Burst & 1.508 & 0.866 & $0.330\pm0.071$ & 293 & 1.28 & 1.0 & 4  \\

J1903+1407 & Null  & -0.325 & 0.900 & $0.432\pm0.040$ & 316 & 2.56 & 2.0  & 13 \\
           & Weak  & 1.280  & 0.870 & $0.374\pm0.032$ & 388 & 1.81 & 1.0  & 10 \\
           & Burst & 3.424  & 1.218 & $0.193\pm0.020$ & 164 & 1.90 & 1.0  & 9  \\
J1502+4653 & Null  & 0.014 & 0.256 & $0.533\pm0.012$ & 385 & 2.81 & 2.0 & 17 \\
           & Weak  & 1.745 & 0.500 & $0.286\pm0.032$ & 351 & 1.76 & 1.0 & 8  \\
           & Burst & 2.578 & 0.589 & $0.181\pm0.032$ & 208 & 1.48 & 1.0 & 9  \\
J2112+4058 & Null  & 0.004 & 0.112 & $0.382\pm0.015$ & 199 & 2.20 & 1.0 & 15 \\
           & Weak  & 1.375 & 0.483 & $0.493\pm0.024$ & 211 & 2.76 & 2.0 & 20 \\
           & Burst & 2.569 & 0.705 & $0.124\pm0.022$ & 92  & 1.16 & 1.0 & 3  \\
\hline
\end{tabular}
\end{table}

\newpage
\bibliography{sample701}{}
\bibliographystyle{aasjournalv7}

\end{document}